\documentclass[journal,twoside,web]{ieeecolor}
\usepackage{color}
\usepackage{generic}
\usepackage{cite}
\usepackage{amsmath,amssymb,amsfonts}
\usepackage{algorithmic}
\usepackage{algorithm}
\usepackage[algo2e]{algorithm2e}
\usepackage{graphicx}
\usepackage{textcomp}
\usepackage{booktabs}
\usepackage{caption}
\usepackage{subcaption}
\def\BibTeX{{\rm B\kern-.05em{\sc i\kern-.025em b}\kern-.08em
    T\kern-.1667em\lower.7ex\hbox{E}\kern-.125emX}}
\newcommand{\bc}{\mbox{\boldmath $c$}}

\newcommand{\bh}{\mbox{\boldmath $h$}}

\newcommand{\br}{\mbox{\boldmath $r$}}
\newcommand{\bq}{\mbox{\boldmath $q$}}
\newcommand{\bs}{\mbox{\boldmath $s$}}

\newcommand{\bx}{\mbox{\boldmath $x$}}
\newcommand{\bw}{\mbox{\boldmath $w$}}

\newcommand{\bC}{\mbox{\boldmath $C$}}

\newcommand{\bH}{\mbox{\boldmath $H$}}
\newcommand{\bI}{\mbox{\boldmath $I$}}

\newcommand{\bS}{\mbox{\boldmath $S$}}
\newcommand{\bU}{\mbox{\boldmath $U$}}
\newcommand{\bV}{\mbox{\boldmath $V$}}
\newcommand{\bW}{\mbox{\boldmath $W$}}
\newcommand{\bQ}{\mbox{\boldmath $Q$}}
\newcommand{\bX}{\mbox{\boldmath $X$}}
\newcommand{\bY}{\mbox{\boldmath $Y$}}
\newcommand{\bZ}{\mbox{\boldmath $Z$}}

\newcommand{\bmu}{\mbox{\boldmath $\mu$}}
\newcommand{\bSigma}{\mbox{\boldmath $\Sigma$}}

\newcommand{\Real}{\mathbb R}

\newcommand{\bzero}{\mbox{\boldmath $0$}}

\newcommand{\be}{\begin{eqnarray}}
\newcommand{\ee}{\end{eqnarray}}

\newtheorem{remark}{Remark}

\begin{document}
\title{Bearing damage detection with orthogonal and non-negative low-rank feature extraction}

\author{Mateusz Gabor, Rafa\l{} Zdunek, Agnieszka Wy\l{}omanska, Rados\l{}aw Zimroz
\thanks{This work is supported by National Center of Science under Sheng2 project No. UMO-2021/40/Q/ST8/00024 "NonGauMech - New methods of processing non-stationary signals (identification, segmentation, extraction, modeling) with non-Gaussian characteristics for the purpose of monitoring complex mechanical structures". 
}
\thanks{M Gabor is with Faculty of Electronics, Photonics, and Microsystems, Wroclaw University of Science and Technology, Wroclaw, Poland, mateusz.gabor@pwr.edu.pl}\thanks{R.Zdunek is with Faculty of Electronics, Photonics, and Microsystems, Wroclaw University of Science and Technology, Wroclaw, Poland,(rafal.zdunek@pwr.edu.pl)}
\thanks{A.Wylomanska is with Faculty of Pure and Applied Mathematics, Hugo Steinhaus Centre, Wroc{\l}aw University of Science and Technology, Janiszewskiego 14a, 50-370 Wroc{\l}aw, Poland (agnieszka.wylomanska@pwr.edu.pl)}
\thanks{R.Zimroz is with Faculty of Geoengineering, Mining and Geology, Wroclaw University of Science and Technology, Na Grobli 15, 50-421 Wroclaw, (radoslaw.zimroz@pwr.edu.pl)}}
\maketitle

\begin{abstract}
Local damage of bearings can be detected as a weak cyclic and impulsive component in a highly noisy measured signal. A key problem is how to extract the signal of interest (SOI) from the raw signal, i.e., how to identify and design an optimal filter. To tackle this problem, we propose to use stochastic sampled orthogonal non-negative matrix factorization for extracting frequency-based features from a spectrogram of the measured signal. The proposed algorithm finds a selective filter that is tailored to the frequency band of the SOI. We show that our approach outperforms the other state-of-the-art selectors that were previously used in condition monitoring. The efficiency of the proposed method is illustrated using both a simulation study and the following real signals: (a) vibration signal from a test rig in the laboratory and (b) acoustic signal from a belt conveyor. 
\end{abstract}


\section{Introduction}
\label{sec:introduction}
Time-frequency representations (TFRs) are frequently used for non-stationary signal processing in machine fault detection \cite{feng2013recent}. One of the most basic, very popular, and intuitive representations is the spectrogram \cite{allen1977short}. It provides a matrix of non-negative values that describe the behavior of the narrowband subsignal energy, varying in time. One reason of using a spectrogram is the easy identification of informative and non-informative frequency bands in a given context (fault detection here). This approach was applied to vibratory surveillance and diagnostics of rotating machines \cite{randall2011rolling} through statistical analysis (spectral kurtosis, SK) of each subsignal (for each frequency bin). Other selectors (various statistics) have also been studied by many research groups  
\cite{hebda2020informative,Chen20229504,Hou2022,Li20205780}. However, considering the spectrogram as a non-negative matrix, non-negative matrix factorization (NMF) has already been used in condition monitoring \cite{wodecki2019novel}. It allows us to identify an informative source and to extract an appropriate time profile or to identify a spectral content at some frequency range. The frequency profile may be simply used as a filter characteristic (after some normalization) to extract the signal of interest (SOI) by multiplying it by the observed signal in the frequency domain. Such an approach requires a selective filter characteristic, i.e., the pass-band for the SOI-related frequencies and the stop-band for other frequencies. After filtering, the SOI should be impulsive and cyclic and should have a better signal-to-noise ratio (SNR) than the raw signal. 

The main purpose of this study is to propose an efficient procedure for fault detection in bearings using a vibration/acoustic signal. As already stated, the detection of damage basically means the detection of a cyclic and impulsive signal in noisy observations. To achieve the goal, the original, raw signal should be pre-filtered in order to enhance the SNR. Practically, the diagnostic information (i.e., an informative part of the raw signal) is located in some frequency bands; however, their location is unknown. 
%

To tackle the filtration problem, we propose a novel approach to fault detection by designing an optimal frequency band (OFB) selector with the orthogonal version of NMF (ONMF), where orthogonality constraints are imposed on the frequency profiles. In this study, we used the stochastic sampled ONMF (SS-ONMF) algorithm that was motivated by the ONMF version proposed by Asteris {\it et al.} \cite{asteris2015orthogonal}. By modifying the selection rule of subspace exploration, the proposed version is more stable for initialization. 

The paper is organized as follows. The problem is introduced in Section \ref{sec:introduction}. Related works on local damage detection are surveyed in Section \ref{sec:related_works}. The basic theory supporting the SOI model, the motivation behind ONMF, and the algorithmic approach are introduced in Section \ref{sec:method}. The proposed procedure is extensively tested in Section \ref{sec:experiments} using advanced simulations, and two real signals from faulty machines with Gaussian and non-Gaussian disturbances. The first real signal is obtained from the test rig in the laboratory, and the second one is taken from the belt conveyor – a heavy-duty machine used in the raw material industry. The second signal is especially interesting, as it also contains some non-Gaussian, impulsive disturbances that make the signal analysis complicated. Finally, our results are compared with the results related to NMF-MU \cite{wodecki2019novel} and popular procedures that are based on a custom filter design for SNR enhancement (spectral kurtosis) \cite{randall2011rolling}, negentropy based information fusion in the concept of infogram \cite{antoni2016info}, conditional variance-based (CVB) selector for non-Gaussian noise \cite{hebda2020informative}, {\color{black} and recently developed sparsogram \cite{zhou2022novel}, which is based on the probabilistic entropy measure}. The paper ends with the conclusions given in Section \ref{sec:conclusions}. 

\section{Related works}
\label{sec:related_works}
There are many techniques, which are based on impulsiveness or periodicity of the SOI (or both)  \cite{antoni2016info}, for detecting local damage in bearings. The impulsive character can be detected by various statistics \cite{Li20205780,Li20221,hebda2020informative}.
To estimate the periodicity of impulses, the envelope spectrum is used \cite{randall2011rolling}. It requires prefiltering before demodulation. The problem of band selection prior to demodulation has been widely discussed in the literature \cite{wang2016new,randall2011rolling, hebda2020informative,Chen20229504,Hou2022,Li20205780}. In addition, Mauricio {\it et al.} \cite{mauricio2020improved} proposed the filter design optimization procedure. The selection of the informative band is especially difficult in the presence of impulsive noise \cite{miao2017improvement_GINI}. 
There are some recent works related to a damage detection problem. Enhanced SVD is proposed by Li {\it et al.} \cite{Li20213220}. They combined E-SVD and the wavelet packet transform to decompose and select an informative band for fault detection. Variational mode decomposition (VMD) based on trigonometric entropy measure was proposed by Kumar {\it et al.} \cite{kumar2021vmd}, where the proposed approach was compared with the classical indicators such as the Kurtosis or Shannon entropy. The novel feature extraction method, called the Shortened Method of Frequencies Selection Nearest Frequency Components (SMOFS-NFC) was proposed for fault classification in electric motors by Głowacz {\it et al.} \cite{glowacz2021fault}. The authors analyzed the acoustic signals from electric impact drills and angle grinders, where after using SMOFS-NFC, the nearest neighbor and naive Bayes classifiers were used. In \cite{glowacz2022thermographic}, the deep neural networks were investigated to thermographic fault diagnosis. Liu {\it et al.} \cite{Liu20221} considered the problem of signal demodulation in the case of non-stationary conditions. Guo {\it et al.} \cite{Guo20221} suggested a combination of a cyclic morphological modulation spectrum and the hierarchical Teager permutation entropy. 
Li {\it et al.} \cite{Li20221} proposed the methodology of modified frequency band envelope kurtosis. Dubey {\it et al.} \cite{Dubey20231} developed an automated variational non-linear chirp mode decomposition while Sun {\it et al.} \cite{Sun2021} proposed a cyclostationary analysis for signals with irregular cyclicity. Furthermore, Sao {\it et al.} \cite{8726382} proposed an impulsive gear fault diagnosis using the methodology based on $\alpha$-stable distribution. \textcolor{black}{Recently, Kumar \textit{et al.} \cite{kumar2022noise} have proposed another approach which is based on the noise subtraction and the marginal enhanced square envelope spectrum for identification of bearing defects in centrifugal and axial pumps, where a priori knowledge on the noise existing in a defect-free bearing is used. The oscillatory behavior-based wavelet decomposition for monitoring the bearing condition in centrifugal pumps was proposed in \cite{kumar2018oscillatory}.} The interesting review of condition monitoring and fault diagnostics based on acoustic signals for different types of faults can be found in \cite{alshorman2021sounds}. 

As the considered signal from a faulty machine is non-stationary and has a time-varying complex spectral structure, a time-frequency representation is often used to understand the properties of the signal and to improve SNR by selecting an informative part. As the time-frequency representation  (TFR) is a matrix of non-negative values (spectrogram here), there is an opportunity to use the NMF philosophy to extract informative components. The family of NMF algorithms is very rich \cite{cichocki2009nonnegative}. Some of them have already been applied to condition monitoring \cite{8942862, wodecki2019novel}. Orthogonal NMF (ONMF) is a special version of NMF, where orthogonality constraints are imposed on one or both of the estimated factors. Such constraints in combination with the non-negativity ones lead to improved sparsity and disjoint features, which is a profitable property in our task. ONMF has been successfully used in many applications, including clustering \cite{10.1145/1150402.1150420,PompiliGAG14,9305974}, facial image representation \cite{4634046}, multispectral document image segmentation \cite{9466419}, hyperspectral image analysis \cite{ctx16793265550003408}, etc. 
To the best of our knowledge, ONMF has never been used to find filter characteristics in a fault detection problem. In our application, we use a stochastic sampled version of orthogonal NMF (ONMF) where the orthogonality constraints are imposed on the frequency profiles to enforce disjoint filter frequency responses. 

\section{Proposed methodology}
\label{sec:method}


{\it Notation:} Matrices, vectors, and scalars are denoted by uppercase boldface letters (e.g., $\bX$), lowercase boldface letters (e.g., $\bx$), and unbolded letters (e.g., $x$), respectively. The set of non-negative real numbers is denoted by $\Real_+$. The $\bx_j$ and $\underline{\bx}_j$ stand for the $j$-th column and row vectors, respectively. Symbols $<\cdot,\cdot>$ and $||\cdot||$ refer to as the inner product and $l_2$ norm. 

\subsection{Model}
\label{sec:model}

Observed signal $y(t)$ is assumed to be a superposition of three components: $s(t)$ -- periodic impulsive SOI, $d(t)$ -- disturbing non-Gaussian (impulsive) signal, and $n(t) \sim \mathcal{N}(0,\sigma^2(f))$ -- disturbing zero-mean Gaussian noise. The latter might be colored by band-pass machine response function $\mathcal{F}_n$  as a result of the transmission path resonance effect. Thus:
\be \label{eq_1} y(t) = s(t) + d(t) + n(t). \ee

The aim is to detect the presence of $s(t)$ in $y(t)$, knowing only that the SOI is a periodic and impulsive signal. This problem can therefore be mathematically formulated in terms of a blind source separation (BSS) problem.     

To justify the proposed methodology and without loss of generality, we assume that signal $s(t)$ can be modeled by a damped sinusoidal signal repeated with period $T_s$. Hence:
\be \label{eq_2} s(t) & = &  A\sum_{m = -\infty}^{\infty} \exp \{ - \alpha_s t\} \sin \left (2 \pi f_s t + \phi_s \right ) \nonumber \\
& \ast &  \delta(t - m T_s),  \ee
where $A$ is the magnitude of the sinusoidal signal, $f_s$ is its frequency, $\alpha_s$ is the damping factor, $\phi_s$ is the initial phase shift, $\delta(\cdot)$ is the Dirac delta function, and symbol $\ast$ is the convolutional operator. We neglected the jitter since it has a marginal effect in our experimental analysis.

Transforming signal $s(t)$ to its time-frequency representation using the short-time Fourier transform (STFT) under the assumption that  window length $M$ satisfies condition: $\frac{1}{2f_s} < M << T_s$ and the spectral leakage effect is neglected, we get spectrogram $\mathcal{S}(f,\tau) = |STFT \{ s(t)\}(f,\tau) |^2$ which may be presented in the factorized form: $\mathcal{S}(f,\tau) = W_s(f) H_s(\tau)$, where
\be \label{eq_3}  W_s(f) = \frac{A^2}{8 \pi \left ( \alpha_s^2 + 4 \pi^2 (f - f_s)^2 \right )} \ee
is a unimodal non-negative spectral function, and $H_s(\tau)$ is a comb function of non-negative period impulses. 

Function $W_s(f)$ determines the frequency response of a band-pass filter with half-power bandwidth $B_s = \frac{\alpha_s \sqrt{\sqrt{2} - 1}}{\pi}$.
 Let  $\mathcal{F}_n(f) = \frac{f_nf}{Q\sqrt{(f_n^2 - f^2)^2 + \frac{f^2 f_n^2}{Q^2}}}$ be the band-pass machine response function (as mentioned above), where $f_n$ is the center frequency and $Q > 0$ is the quality factor. The bandwidth of $\mathcal{F}_n(f)$ amounts to $B_n = f_n \left ( \frac{2Q +1 - \sqrt{1 + 4Q^2}}{Q}\right )$.
Assuming
\be \label{eq_4} 
\left | (f_s \pm \frac{B_s}{2}) - (f_n \pm \frac{B_n}{2}) \right | > \epsilon,
\ee
where $\epsilon > 0$ is some frequency separation threshold and neglecting the influence of signal $d(t)$, one can notice that $\left < W_s(f), \mathcal{F}_n(f) \right >_{\epsilon} = \lim_{\epsilon \rightarrow \infty}\int_{-\infty}^{\infty} W_s(f) \mathcal{F}_n(f) df \rightarrow 0$, which means that functions $W_s(f)$ and $\mathcal{F}_n(f)$ are orthogonal in the entire frequency domain.   

Unfortunately, the above-mentioned orthogonality condition may be slightly disturbed in practice by considering the following circumstances: (1) $d(t)$ is a broadband impulsive noise that may also fall into the pass-band of $W_s(f)$ -- fortunately, only for relatively few time instances, (2) the energy of $n(t)$ is usually much higher than the energy of $s(t)$, and despite the action of $\mathcal{F}_n(f)$, some part of the energy of $n(t)$ may be located in the pass-band of $W_s(f)$. Thus, functions $W_s(f)$ and $\mathcal{F}_n(f)$ are only quasi-orthogonal in practice, which motivates the use of the ONMF model that provides a chain of stochastic approximations to the above-mentioned BSS problem. 

\subsection{NMF}
\label{sec:NMF}
Let $\bY \in \Real_+^{I \times T}$ be a discretized version of spectrogram $\mathcal{Y}(f,\tau)$ of $y(t)$. The discretized version of $\mathcal{S}(f,\tau)$ can be represented in the factorized form: $\bS  = \bw_s \bh^T_s \in \Real_+^{I \times T}$, where $\bw_s \in \Real_+^I$ and $\bh_s \in \Real_+^T$ represent $W_s(f)$ and $H_s(\tau)$, respectively. Similarly, the spectrograms of $d(t)$ and $n(t)$ can also be expressed by the rank-1 factorization models with frequency profiles $\bw_d \in \Real_+^I$ and $\bw_n \in \Real_+^I$, and time profiles $\bh_d \in \Real_+^T$ and $\bh_n \in \Real_+^T$. 
Assuming the orthogonality conditions for any pair of frequency profiles, i.e., $\bw_s^T \bw_d = 0$, $\bw_s^T \bw_n = 0$, and $\bw_d^T \bw_n = 0$, spectrogram $\bY$ of model (\ref{eq_1}) can be expressed by the ONMF model: 
\be \label{eq_5} 
\bY & = & \bW \bH^T, \quad {\rm where} \quad \bW^T\bW = \bI_R, \nonumber \\
& & \bW = [ \bw_s, \bw_d, \bw_n] \in \Real_+^{I \times R}, \; \bH \in \Real_+^{T \times R}, 
\ee
where $\bI_R \in \Real^{R \times R}$ is an identity matrix, and $R = 3$ for the discussed case.  

It is well known that the orthogonality constraint combined with the non-negativity one leads to an increase in sparsity of the estimated profiles. This is a highly desirable property for both profiles. The sparse and mutually disjoint frequency profiles determine the magnitude frequency responses of the band filters. Due to the non-negativity and orthogonality constraints imposed onto the columns of $\bW$,  we have $\bH = \bY^T \bW$, and hence, each row vector of $\bY$ can be regarded as a scaled version of the selected column of $\bH$. If SOI is observed in the spectrogram of $y(t)$ even in a narrowband, then this information should be reflected in the right time profile in $\bH$ due to the selective filtration imposed by $\bW$. The orthogonality constraints allow us to select the time profiles as the most probable row vectors of $\bY$, and the frequency profiles as the magnitude frequency responses of the adaptive filters. 

Using the Euclidean distance measure and considering the orthogonality constraints, the ONMF problem can be equivalently reformulated by the following maximization problem: 
\be \label{eq_6} \max_{{\bf W} \in \mathcal{W}} ||\bY^T \bW ||_F^2, \ee where 
\be \label{eq_7} \mathcal{W} = \left \{\bW \in \Real_+^{I \times R}, \quad \bW^T\bW = \bI_R \right \}. \ee

Note that problem (\ref{eq_6}) searches for feature vectors in $\bW$ that maximize the energy of the components in $\bH$. Without the non-negativity constraints, problem (\ref{eq_6}) would express the standard principal component analysis (PCA) problem. Considering the non-negativity and orthogonal constraints, it becomes the non-negative PCA (NPCA) problem. 

\subsection{Algorithmic approach}
\label{sec:algorithm}

Problem (\ref{eq_6}) is non-convex and due to the non-negativity constraints, it is also NP-hard. However, Asteris {\it et al.} \cite{asteris2015orthogonal} noticed that it can be considerably simplified by a combinatorial approximation of the low-rank subspace to  input matrix $\bY$. Let $\Psi = ||\bY^T \bW||_F^2$ be the objective function in (\ref{eq_6}). Applying the SVD to $\bY$, we have: $\bY = \bU\bSigma\bV^T$, and $\Psi = ||\bV \bSigma^T \bU^T \bW||_F^2 = ||\bSigma^T \bU^T \bW||_F^2$ since $\bV^T\bV = \bI$. Let $\tilde{\bY} = {\it LR}^{(J)}(\bY) = \tilde{\bU} \tilde{\bSigma} \tilde{\bV}^T \in \Real^{I \times T}$ be the rank-$J$ approximation to $\bY$, where $J < {\it rank}_{\epsilon}(\bY)$, and $\tilde{\bU} \in \Real^{I \times J}$, $\tilde{\bSigma} = {\rm diag}(\sigma_j) \in \Real^{J \times J}$, and $\tilde{\bV} \in \Real^{T \times J}$. Therefore, the maximization problem in (\ref{eq_6}) can be reduced to the low-rank approximation problem: 
\be \label{eq_8} \max_{{\bf W} \in \mathcal{W}} ||\bZ \bW ||_F^2, \ee
where $\bZ = \tilde{\bSigma} \tilde{\bU}^T \in \Real^{J \times I}$. 

Problem (\ref{eq_8}) can be further simplified by efficient sampling of the column space of $\tilde{\bY}$. The set of candidate solutions can be determined by the $R$-dimensional $l_2$-unit sphere $\mathcal{S}^{R-1}$. Using the Cauchy-Schwartz inequality, the objective function in (\ref{eq_8}) can be approximated by: 
\be \label{eq_9} \tilde{\Psi}  & = &  ||\tilde{\bSigma} \tilde{\bU}^T \bW ||_F^2 
= \sum_{r = 1}^R ||\bZ \bw_r||^2 = \sum_{r = 1}^R ||\bZ \bw_r||^2 ||\bc_r||^2 \nonumber \\ & \geq & \sum_{r = 1}^R \left < \bZ\bw_r,\bc_r \right >^2, \quad \forall \bc_r \in \Real^J, \quad ||\bc_r|| = 1. 
\ee
Taking into account (\ref{eq_9}), problem (\ref{eq_8}) can be rewritten to the form: 
\be \label{eq_9a} \max_{{\bf W} \in \mathcal{W}} \sum_{r = 1}^R \max_{\bc_r \in  \mathcal{S}^{R-1}} \left <\bZ\bw_r,\bc_r \right >^2. \ee

The solution to (\ref{eq_9a}) can be found by applying the concept of alternating optimization, switching between the updates for $\{ \bc_r\}$ and $\bW$. However,  this approach would not be more effective in terms of computational complexity than solving directly problem (\ref{eq_8}). An alternative solution was proposed by Asteris {\it et al.} \cite{asteris2015orthogonal} who used a stochastic sampling procedure to select $\bc_j$ from a finite $\epsilon$-net on $\mathcal{S}^{R-1}$, and then to estimate $\bW$ given the sample $\bc_r$. This net, referred to as $\mathcal{N}_{\epsilon}^{\otimes k}(\mathcal{S}^{R-1})$, determines the current sampling point within distance $\epsilon$ from the former. This concept is justified by the fact that the second problem in (\ref{eq_9a}) is maximized when $\forall r: \bc_r$ is collinear with $\bZ\bw_r$. 

To estimate $\bw_r$ given $\bc_r$ for $r = 1, \ldots, R$ in (\ref{eq_9a}), the objective function can be rewritten as follows: $\left <\bZ\bw_r,\bc_r \right >^2 = \left <\bZ^T \bc_r, \bw_r \right >^2$, which leads to the problem: 
\be \label{eq11} \max_{{\bf W} \in \mathcal{W}} \sum_{r = 1}^R \left <\bq_r, \bw_r \right >^2 \ee
where $\bq_r = \bZ^T \bc_r$ and $\mathcal{W}$ is defined in (\ref{eq_7}). Problem (\ref{eq11}) can be efficiently solved by finding an optimal support for an unknown vector $\bw_r$, that is, the set that contains the indices of active entries. Since the objective function is expressed by the inner product $\left <\bq_r, \bw_r \right >$ and $\bq_r$ is unconstrained, the optimal support may contain the indices of $\bq_r$ that correspond only to non-negative or non-positive entries, excluding their combination. Obviously, zero-value entries are not active variables and do not affect the objective function. 

Asteris {\it et al.} \cite{asteris2015orthogonal} proposed exploring the space of optimal supports by taking all $2^R$ possible sign combinations for the support sets. As a result, problem (\ref{eq11}) in \cite{asteris2015orthogonal} is solved on all scaled matrices $\tilde{\bQ} = \bQ {\rm diag} \{ \bs\}$, where $\bs \in \{ \pm 1\}^R$. 
Let $\Gamma_r = \{i: q_{ir} > 0 \}$ be the optimal support for $r = 1, \ldots, R$. Considering the Cauchy-Schwartz inequality:
\be \label{eq12} \left <\bq_r, \bw_r \right >^2 & = & \left <\bq_r({\Gamma_r}), \bw_r({\Gamma_r}) \right >^2  \leq  ||\bq_r({\Gamma_r})||^2 ||\bw_r({\Gamma_r})||^2 \nonumber \\ & = & ||\bq_r({\Gamma_r})||^2, \ee
the solution to (\ref{eq11}) takes the form: 
\be \label{eq13} \bw_r({\Gamma_r}) = \frac{\bq_r({\Gamma_r})}{||\bq_r({\Gamma_r})||} \quad {\rm for} \quad \bq_r({\Gamma_r}) \geq 0. \ee
Thus: $\left <\bq_r, \bw_r \right >^2 = \sum_{i \in \Gamma_r} q_{ir}^2$. 

 Finally, the orthogonal NMF through subspace exploration (ONMFS) \cite{asteris2015orthogonal} can be formalized with Algorithm \ref{alg1}. 
 The update rule for sampling candidate matrix $\bW_c$ is presented in the form of Algorithm \ref{alg2}, and it is referred to as {\tt UpdateW \textunderscore onmfs}.
 
\begin{algorithm}[h!]
\caption{ \bf ONMFS} \label{alg1}
\SetKwInOut{Input}{Input} \SetKwInOut{Output}{Output}
\Input{$\bY \in \Real_+^{I \times T}$ -- spectrogram of input signal $y(t)$, \\ $J$ -- rank of approximation, \\ $R$ -- number of extracted filters, \\ $K$ -- maximum number of iterations}
\Output{$\bW \in \Real_+^{I \times R}$ -- orthogonal matrix of filters, \\ $\bH$ -- time-profiles}
 \BlankLine
 $\left [\tilde{\bU},\tilde{\bSigma},\tilde{\bV} \right ]  = {\tt svd}(\bY,J)$\tcp*[f]{Truncated SVD} \\
 $\bZ = \tilde{\bU}\tilde{\bSigma} \in \Real^{I \times J}$, $\mathcal{C} = \emptyset$, \\
 \ForEach{$\bC \in \mathcal{N}_{\epsilon}^{\otimes k}(\mathcal{S}^{R-1})$ }{
 $\bQ = \bZ\bC \in \Real^{I \times R}$, \\
 $\bW_c = {\tt UpdateW \textunderscore onmfs}(\bQ)$\tcp*[f]{Algorithm \ref{alg1}} \\
 $\mathcal{C} \leftarrow \mathcal{C} \; \bigcup \; \{ \bW_c \}$, 
 }
 $\bW_* = \arg \max_{{\bf W} \in \mathcal{C}} ||\bZ^T \bW||_F^2$, \\
 $\bH = \bY^T\bW_* \in \Real_+^{T \times R}$, 

 \end{algorithm}

 \begin{algorithm}[h!] 
\caption{\bf UpdateW\textunderscore onmfs} \label{alg2}

\SetKwInOut{Input}{Input} \SetKwInOut{Output}{Output}
\Input{$\bQ = [q_{ir}] \in \Real_+^{I \times R}$ -- sampled orthogonal components}
\Output{$\bW \in \Real_+^{I \times R}$ -- candidate matrix $\bW$ }
 \BlankLine
  $\mathcal{C}_W = \emptyset$, \\
 \ForEach{$\bs \in \{ \pm 1\}^R$ }{
 $\tilde{\bQ} = \bQ {\rm diag} \{ \bs\}$, \\
 $\Gamma_r = \emptyset$ for $r = 1, \ldots, R$, \\
 \For{$i = 1, \ldots, I$}{
 $r_* = \arg \max_r \tilde{q}_{ir}$, \\
 \If{$\tilde{q}_{i,r_*} \geq 0$}
 {
 $\Gamma_{r_*} \leftarrow \Gamma_{r_*} \; \bigcup \; \{ i\} $, 
 }
 }
 $\tilde{\bW} = \bzero \in \Real^{I \times R}$, \\
 \For{$r = 1, \ldots, R$}
 {
 $\tilde{\bw}_r(\Gamma_r) = \frac{\tilde{\bq}_r(\Gamma_r)}{||\tilde{\bq}_r(\Gamma_r)||}$, 
 }
 $\mathcal{C}_W \leftarrow \mathcal{C}_W \; \bigcup \; \tilde{\bW}$, \; {\rm where} \; $\tilde{\bW} = [\tilde{\bw}_1, \ldots, \tilde{\bw}_R]$,
 }
 $\bW = \arg \max_{{\bf W} \in \mathcal{C}_W} \sum_{r = 1}^R \left <\bq_r, \bw_r \right >^2$, 

 \end{algorithm}

Net $\mathcal{N}_{\epsilon}^{\otimes k}(\mathcal{S}^{R-1})$ in Algorithm \ref{alg1} is modeled with the zero-mean Gaussian distribution, i.e. $\bc_r \sim \mathcal{N}(0,1)$, which guarantees independence and symmetry of sampling. However, this independence and the fixed parameters of a reference distribution (from which samples are drawn) result in a long chain of samples. In this large set, only a few samples may satisfy the selection criterion, which is computationally ineffective and may lead to a higher sensitivity of the whole  algorithm to random initialization.  

To tackle the above-mentioned problems, we propose to determine a finite $\epsilon$-net on the Markov chain combined with an evolutionary strategy where the variance of samples, which are generated from the reference conditional distribution, changes with iterations. Candidate sample $\hat{\bc}_r^{(k+1)}$ is generated from conditional distribution $P(\bc_r|\bc_r^{(k)})$ in the $k$-th iteration, where $\bc_r^{(k)}$ is the sample of current state $k$. Sample $\hat{\bc}_r^{(k+1)}$ is accepted, that is, $\bc_r^{(k+1)} = \hat{\bc}_r^{(k+1)}$, if $\Psi(\hat{\bc}_r^{(k+1)}) > \Psi(\bc_r^{(k)})$. Thus: $\forall r: P(\bc_r^{(k+1)}| \bc_r^{(k)}, \ldots, \bc_r^{(0)}) = P(\bc_r^{(k+1)}| {\bc}_r^{(k)})$, which determines the Markov chain. We set $P(\bc_r|\bc_r^{(k)}) = \mathcal{L}(\bc_r^{(k)},\beta^{(k)} {\bold e})$, where $\mathcal{L}$ is the multivariate isotropic Laplacian distribution with mean $\bmu = \bc_r^{(k)}$ and scale parameter $\beta^{(k)}$. This choice was motivated mostly by experimental observations. With reference to the Gaussian distribution, the Laplacian one is more suitable for sparse components, i.e., it has higher values of the probability density function around the mean and a longer tail. As a result, only a few entries of $\bc_r^{(k)}$ are strongly perturbed, and the others are given to weaker perturbations. The variance of the perturbations is also controlled by scale parameter $\beta^{(k)}$. We assumed the following rule: 
\be \label{eq10} \beta^{(k)} = \max \{ \bar{\epsilon}, 1 - {\rm tanh(k)} \}, \ee
where $1 >> \bar{\epsilon} > 0$ protects against reaching zero variance. The function in rule (\ref{eq10}) behaves similarly to an inverse logistic function. For $k = 0$, $\beta^{(k)} = 1$, and it asymptotically decreases with $k$ to $\bar{\epsilon}$. Regarding the process of generating the samples in the context of the evolutionary strategy, rule (\ref{eq10}) favors the exploration phase in the early steps because the variance of the candidate samples is large. When the number of iterations is sufficiently large, the variance is small, and it gives privilege to the exploitation phase. 


Furthermore, Algorithm \ref{alg1} suffers from a large instability of the output factors because candidates $\{ \bW_c\}$ are generated with a stochastic sampling rule, and the orthogonality and nonnegativity constraints are not sufficient to guarantee each $\bW_c$ is a full-rank matrix. To tackle this problem, we changed the selection rule to accept only the candidates whose rank is greater than one.  However, this assumption is still very weak in the discussed application. It is easy to note that matrix $\bW \in \Real_+^{I \times R}$ with $\bW^T \bW = \bI_R$, where one its column has only one positive entry and zero-entries otherwise, satisfies all the mentioned constraints (non-negativity, orthogonality, and full-rank). Unfortunately, such a candidate would generate a terrible narrow-band filter, useless for the selection of the SOI. Considering our prior knowledge on the spectral properties of the SOI, we may roughly estimate lower and upper bounds for the bandwidth of the desired filter. To incorporate the bounds, the $l_0$-norm is calculated for each column in $\bW_c$. Then we assumed that $\forall r: ||\bw_r||_0 > \xi I$, where $\xi > 0$ is the user-defined parameter controlling the lower-bound.  
To determine the upper-bound, we presumed that the spectral properties of the noisy perturbance and the SOI are considerably different, and hence, the corresponding filters cannot have similar bandwidths. The non-uniformity in the spectral characteristics is enforced by the assumption $\max_r(l_r) - \min_r(l_r) > {\rm mean}(l_r)$, where $l_r$ is the $l_0$-norm of the $r$-th column vector of $\bW_c$. 

The final version of the SS-ONMF algorithm, in which matrix $\bW$ is estimated by solving problem (\ref{eq_6}), is given by Algorithm \ref{alg3}. The function {\tt UpdateW} is listed by Algorithm \ref{alg4}. The input to this algorithm is unconstrained matrix $\bQ \in \Real^{I \times R}$, and the output is non-negatively constrained orthogonal matrix $\bW \in \Real_+^{I \times R}$ and objective function $\Psi$ in (\ref{eq11}). The algorithm yields a set of optimal supports $\Gamma_r$ for each column of $\bQ$. Each $\Gamma_r$ contains the indices of the rows for which $q_{ir}$ is non-negative and the largest among the elements in this row.

\begin{algorithm}[h!]
\caption{\bf SS-ONMF} \label{alg3}
\SetKwInOut{Input}{Input} \SetKwInOut{Output}{Output}
\Input{$\bY \in \Real_+^{I \times T}$ -- spectrogram of input signal $y(t)$, \\ $J$ -- rank of approximation, \\ $R$ -- number of extracted filters, \\ $K$ -- maximum number of iterations, \\  $\xi$ -- minimum bandwidth factor}
\Output{$\bW \in \Real_+^{I \times R}$ -- orthogonal matrix of filters, \\ $\bH$ -- time-profiles}
 \BlankLine
 $\left [\tilde{\bU},\tilde{\bSigma},\tilde{\bV} \right ]  = {\tt svd}(\bY,J)$\tcp*[f]{Truncated SVD} \\
 $\bZ = \tilde{\bU}\tilde{\bSigma} \in \Real^{I \times J}$, $\bC = \bzero \in \Real^{J \times R}$, $\Psi = 0$, \\
 \For{$k = 1, \ldots, K$}{
 $\beta^{(k)}$ given by (\ref{eq10}), \\
 $\bC^{(k)} = \bC + \beta^{(k)} \mathcal{L}(\bzero,{\bf e})$, \\
 $\forall r: \bc_r^{(k)} \leftarrow \frac{\bc_r^{(k)}}{||\bc_r^{(k)}||}$\tcp*[f]{Proj. onto sphere $\mathcal{S}^{R-1}$} \\
 $\bQ^{(k)} = \bZ\bC^{(k)} \in \Real^{I \times R}$, \\
 $\left [\bW^{(k)}, \Psi^{(k)} \right ] = {\tt UpdateW}(\bQ^{(k)})$\tcp*[f]{Algorithm \ref{alg3}} \\
 $\forall r: l_r^{(k)} = ||\bw_r^{(k)}||_0$, \tcp*[f]{$l_0$ norm} \\
 \If{$\Psi^{(k)} > \Psi$ \; $\wedge$ \; ${\rm rank}(\bW^{(k)}) > 1$ \;
 $\wedge$ \; $\forall r: l_r^{(k)} > \xi I$, \; $\wedge$ \; $\max_r(l_r^{(k)}) - \min_r(l_r^{(k)}) > {\rm mean}(l_r^{(k)})$
 }
 {
 $\Psi = \Psi^{(k)}$, $\bW = \bW^{(k)}$, $\bC = \bC^{(k)}$, 
 }
 }
 $\bH = \bY^T\bW \in \Real_+^{T \times R}$
 \end{algorithm}

\begin{algorithm}[h!]
\caption{\bf UpdateW} \label{alg4}
\SetKwInOut{Input}{Input} \SetKwInOut{Output}{Output}
\Input{$\bQ = [q_{ir}] \in \Real_+^{I \times R}$ -- sampled orthogonal components}
\Output{$\bW \in \Real_+^{I \times R}$ -- candidate matrix $\bW$, \\
$\Psi$ -- objective function}
 \BlankLine
 $\bar{\br} = [\bar{r}_i] \in \Real^I$, where $\bar{r}_i = \arg \max_r q_{ir}$, \tcp*[f]{Indices} \\
 $\bar{\bq} = [\bar{q}_i] = [q_{i,\bar{r}_i}] \in \Real^I$,\tcp*[f]{Maximum values}  \\
 $\bW = \bzero \in \Real^{I \times R}$, \\
 $\Psi = \sum_{i_* \in \Gamma_+} \bar{q}_{i_*}^2$, \; {\rm where} \; $\Gamma_+ = \{i: \bar{q}_i > 0 \}$, \tcp*[f]{Obj.} \\
 \For{$r = 1, \ldots, R$}
 {
 $\Gamma_r = \{i: \bar{r}_i = r \; \wedge \; \bar{q}_i \geq 0 \}$, \\
 $\bw_r(\Gamma_r) = \displaystyle{ \frac{\bar{\bq}(\Gamma_r)}{||\bar{\bq}(\Gamma_r)||} }$,
 }
 \end{algorithm}

\begin{remark}
\label{r1}
The computational complexity of Algorithm \ref{alg3} is strongly affected by the time complexity of the truncated SVD (TSVD) algorithm. Assuming $I < T$, and the QR factorization combined with the economy-size SVD is used to compute the TSVD of $\bY$, the time complexity of the TSVD can be roughly estimated as $\mathcal{O}(2I^2T + I^3 + IJT)$. In the loop {\bf for}, the most computational task is the computation of matrix $\bQ^{(k)}$, which needs $\mathcal{O}(IJRK)$, where $K$ is a number of iterations. The time complexities of other steps, including Algorithm \ref{alg4}, can be neglected. Thus, the overall computational complexity of Algorithm \ref{alg3} becomes dominated by the computational operations in the loop {\bf for}, if $K > \frac{2IT}{JR}$, which usually occurs for our datasets if $K$ exceeds $10^4$ iterations.  
\end{remark}

\section{\textcolor{black}{Experimental setup}}
\label{sec:experiments}

\begin{figure}[H]
    \centering
    \includegraphics[scale=0.17]{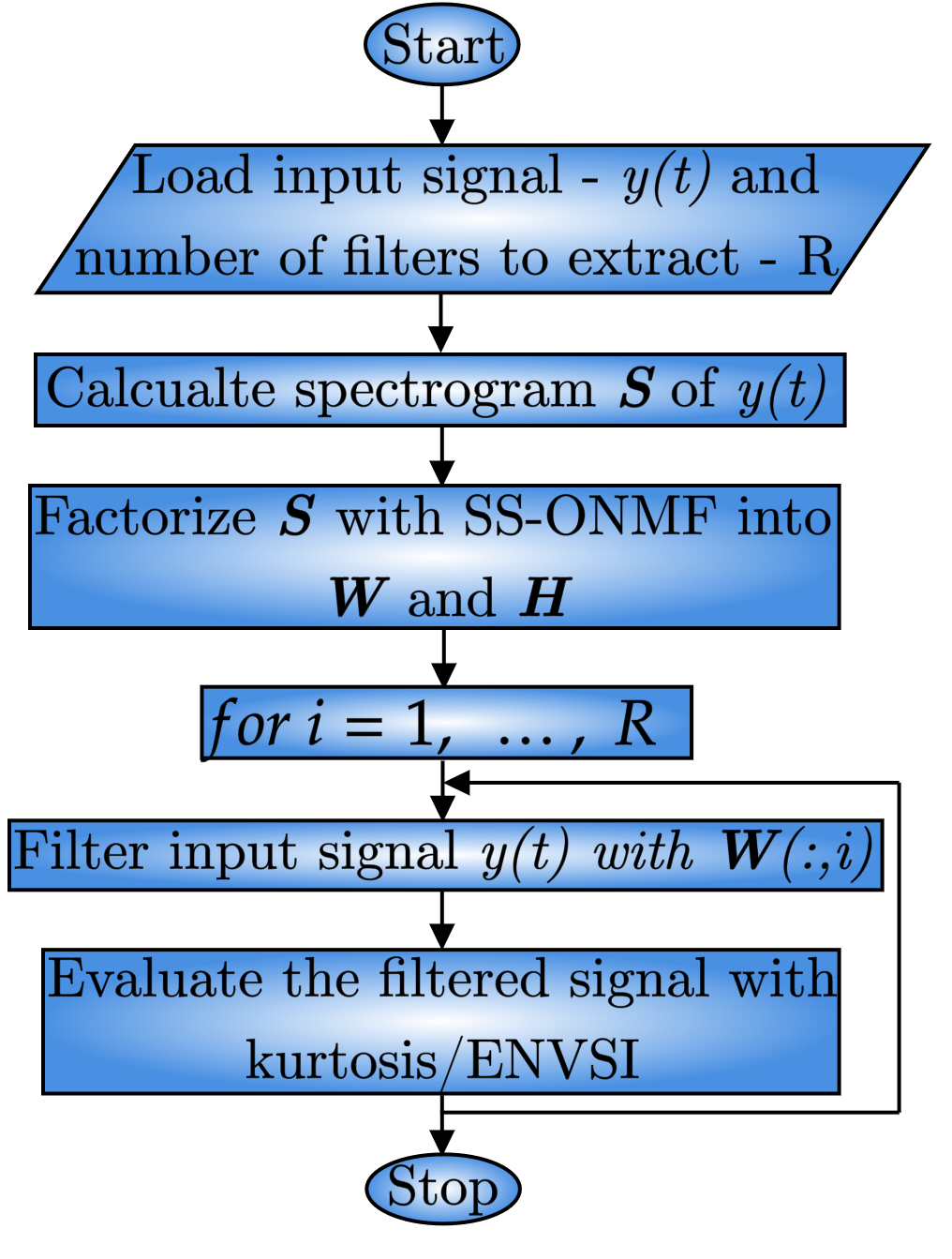}
    \caption{The flowchart of proposed method.}
    \label{fig:flowchart}
\end{figure}

The input to the experiments is the mixed signal of sources (the SOI plus noisy components), which is transformed into a spectrogram, and then decomposed into $\bW$ and $\bH$ matrices using the SS-ONMF method. Subsequently, matrix $\bW$ consists of $R$ OFB selectors used to filter the original signal. This procedure consists of five steps: 1) computation of the spectrogram of the diagnostic signal, 2) factorization of the spectrogram, 3) selection of an appropriate band, 4) filtering of the original signal, 5) evaluation of impulsiveness (by kurtosis) or periodicity (estimation of the envelope-based spectral indicator (ENVSI) \cite{hebda2020informative}). \textcolor{black}{We remind that the kurtosis is considered as the most popular measure of impulsiveness, and thus it can be applied for evaluation of the received results (to indicate the impulsive behavior of the signal) in the case with Gaussian noise.  Moreover, the ENVSI is applied to demonstrate the periodicity. It is well-known that the kurtosis is sensitive to large impulses. Thus, in the non-Gaussian scenario, we use only the ENVSI for evaluation of the results. } The flowchart of the procedure is presented in Figure \ref{fig:flowchart}. 

The proposed SS-ONMF algorithm is compared with the following baseline and state-of-the-art OFB selectors: orthogonal NMF through subspace exploration (ONMFS)\footnote{The Matlab code of ONMFS was downloaded from website: https://github.com/megasthenis/spanonmf} \cite{asteris2015orthogonal}, NMF with multiplicative updates (NMF-MU) \cite{wodecki2019novel}, SK \cite{randall2011rolling}, CVB \cite{hebda2020informative}, infogram \cite{antoni2016info}, {\color{black} and sparsogram \cite{zhou2022novel}}. The NMF-based algorithms were tested for the ranks in the range $[6, 15]$. Due to the non-convexity problem in NMF algorithms, 100 Monte Carlo (MC) trails were performed for each rank in all NMF methods. To fairly select optimal filters, we calculated the medians of kurtosis/ENVSI values of filtered signals for Gaussian/non-Gaussian noise for each rank from the set generated with MC simulations. The median is chosen because the results of kurtosis/ENVSI demonstrate a skewed distribution. The best rank is the rank for which the median of kurtosis/ENVSI is the highest.

All the spectrograms were computed using the Hamming window of the length equal to 128, the overlap set to 100 samples, and the number of discrete Fourier transform (DFT) points set to 512. 

The experiments are performed on two types of signals: simulated and real ones. For the simulated signals, it will be shown how, on average, the resulting filter obtained by SS-ONMF is resistant to Gaussian and non-Gaussian noise. In the experiments with real signals, the robustness of SS-ONMF will be analyzed with respect to ONMFS, and the obtained filters will be compared with other OFB selectors. 

\subsection{Simulated signals}
\label{sub:simulated_signals}

\subsubsection{Gaussian noise}
\label{subsub:gaussian_signals}

\begin{figure}[h!]
    \centering

    \centering
    \begin{subfigure}[b]{0.49\linewidth}
         \centering
         \includegraphics[width=\linewidth]{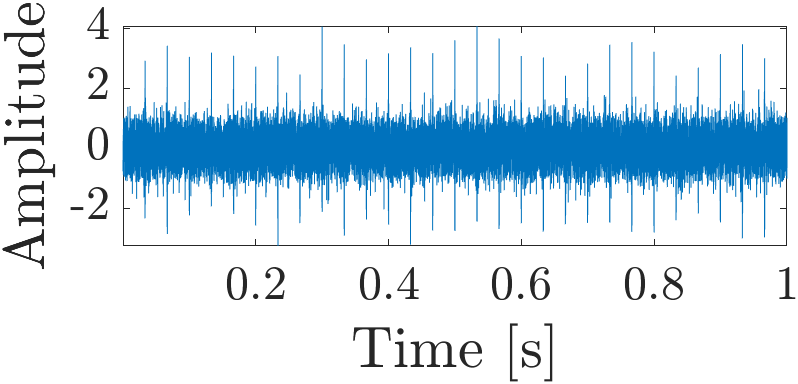}
         \caption{$\sigma=0.5$}
     \end{subfigure}
     \hfill
     \begin{subfigure}[b]{0.49\linewidth}
         \centering
         \includegraphics[width=\linewidth]{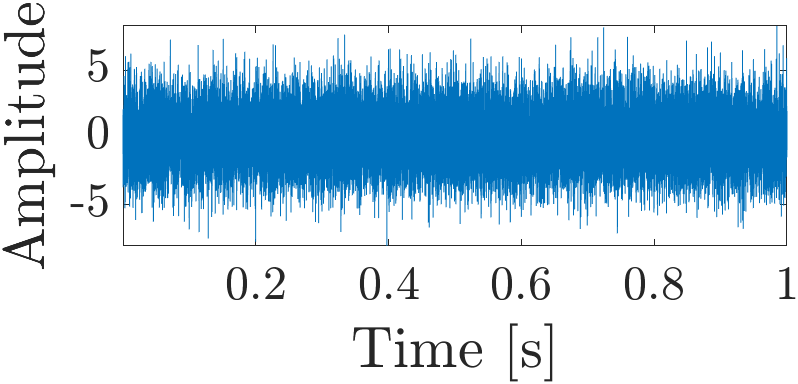}
         \caption{$\sigma=2.0$}
     \end{subfigure}
     \vskip\baselineskip
     \begin{subfigure}[b]{0.49\linewidth}
         \centering
         \includegraphics[width=\linewidth]{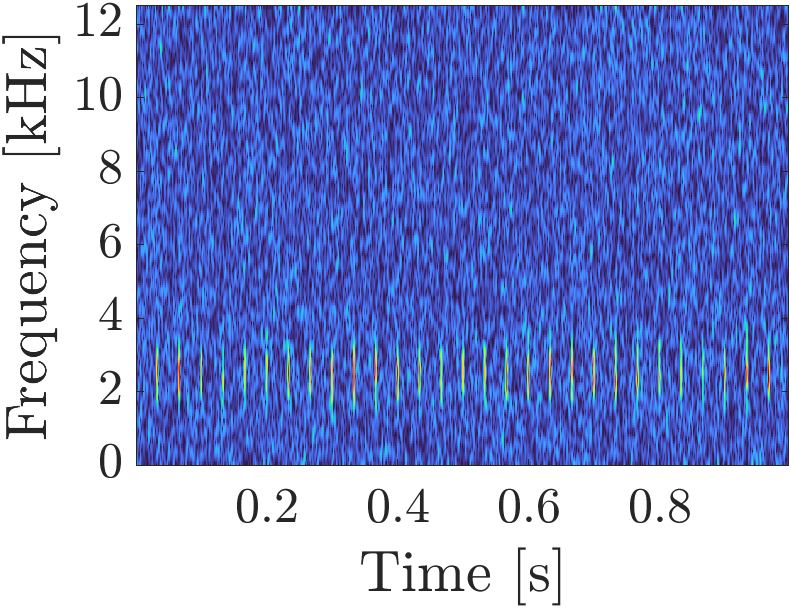}
         \caption{$\sigma=0.5$}
     \end{subfigure}
     \hfill
     \begin{subfigure}[b]{0.49\linewidth}
         \centering
         \includegraphics[width=\linewidth]{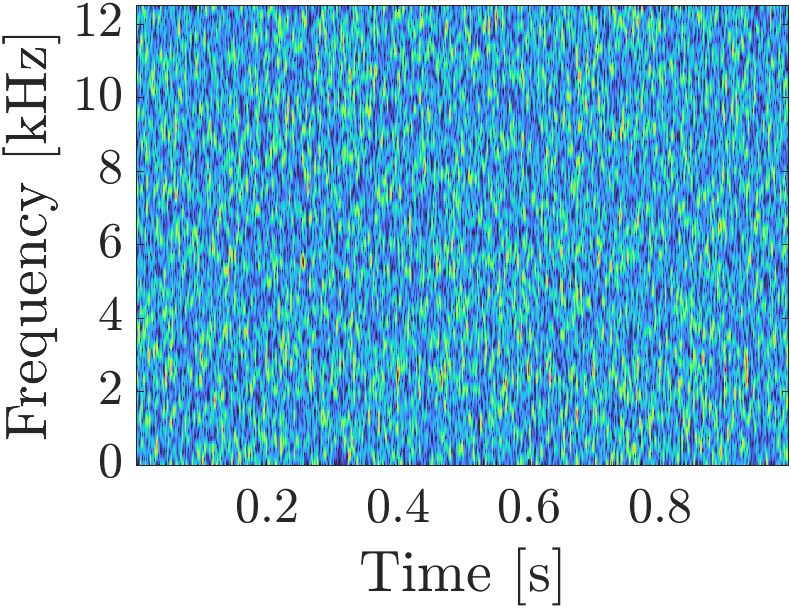}
         \caption{$\sigma=2.0$}
     \end{subfigure}

    
    \caption{Simulated signals with Gaussian noise and their spectrograms: (a), (c) Gaussian noise with $\sigma=0.5$, (b), (d) Gaussian noise with $\sigma=2.0$.}
    \label{fig:SOI_gauss}
\end{figure}

We generated three signals with cyclic impulses of the amplitude equal to 3 and the Gaussian noise with $\sigma$ (standard deviation) equal to: 0.5, 1.7, and 2.0. For the weakest noise, the cyclic impulses are clearly visible over the noisy background (easy case), and they are completely masked by the noise when the standard deviation is the largest (see Figure \ref{fig:SOI_gauss}). The fault frequency is 30 Hz, and the carrier frequency of the cyclic impulses is 2.5 kHz (center of OFB).

\subsubsection{Non-Gaussian noise}
\label{subsub:nongaussian_signals}
\begin{figure}[h!]

    \centering
    \begin{subfigure}[b]{0.49\linewidth}
         \centering
         \includegraphics[width=\linewidth]{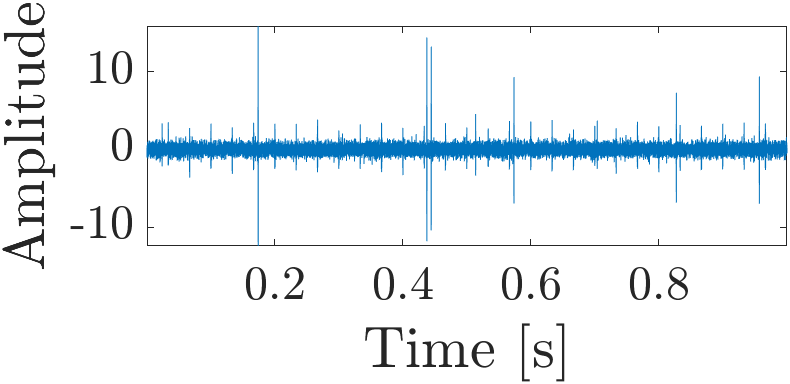}
         \caption{$\sigma=0.5$}
     \end{subfigure}
     \hfill
     \begin{subfigure}[b]{0.49\linewidth}
         \centering
         \includegraphics[width=\linewidth]{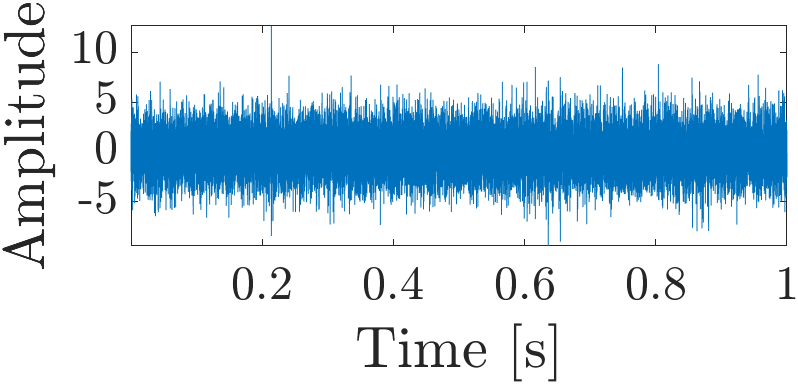}
         \caption{$\sigma=2.0$}
     \end{subfigure}
     \vskip\baselineskip
     \begin{subfigure}[b]{0.49\linewidth}
         \centering
         \includegraphics[width=\linewidth]{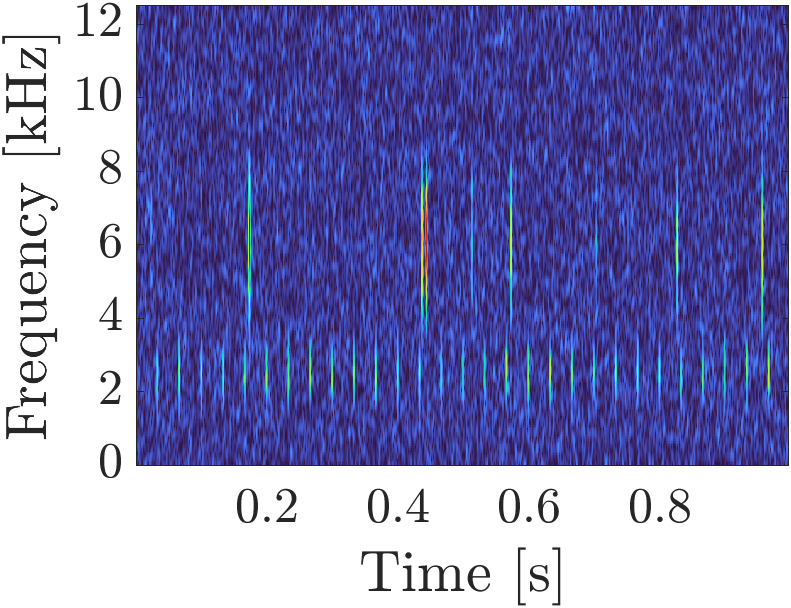}
         \caption{$\sigma=0.5$}
     \end{subfigure}
     \hfill
     \begin{subfigure}[b]{0.49\linewidth}
         \centering
         \includegraphics[width=\linewidth]{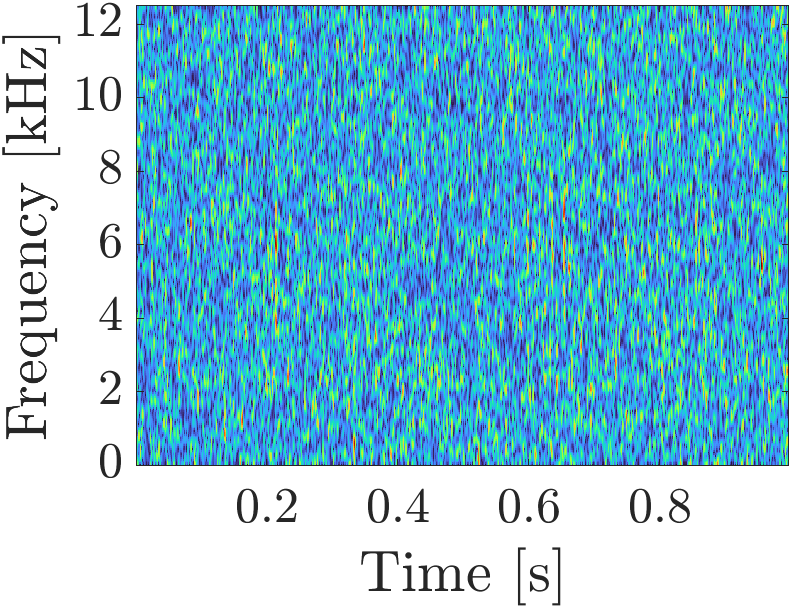}
         \caption{$\sigma=2.0$}
     \end{subfigure}

    \caption{Simulated signals with non-Gaussian noise of max. amplitude of non-Gaussian impulses equal to 15: (a), (c) Gaussian noise with $\sigma=0.5$, (b), (d) Gaussian noise with $\sigma=2.0$.}
    \label{fig:SOI_nonGauss}
\end{figure}

The non-Gaussian noise considered here is a mixture of Gaussian noise and additive outliers (non-Gaussian impulses) with a given amplitude. In consequence, the noise has a non-Gaussian distribution with a possible large tail. In the simulation study, six signals are generated with cyclic impulses of the amplitude equal to 3, the Gaussian noise with $\sigma$ taking the values: 0.5, 1.1, 2.0, and the non-Gaussian impulses with the amplitudes: 5 and 15. The fault frequency is 30 Hz, the carrier frequency of the cyclic impulses is 2.5 kHz, and the carrier frequency of non-Gaussian impulses is 6 kHz. Exemplary signals and their spectrograms are presented in Figure \ref{fig:SOI_nonGauss}.

\subsection{Real signals}
\label{sub:real_signals}

Similarly to simulations, we selected two real signals -- the ones disturbed by only Gaussian noise, and the other containing Gaussian noise with some extra wideband impulses. \textcolor{black}{The parameters of bearings used in the experiments are presented in Table \ref{tab:bearings}.}

\begin{table}[]
\centering
\caption{\textcolor{black}{Bearings used in experiments and their parameters. The vibration signal was obtained from SKF 1205 EKTN9 bearing, and the acoustic signal was measured on SKF 6204-2RSH bearing.}}
\label{tab:bearings}
\begin{tabular}{cccccc}
\hline
\textbf{Bearing} & \textbf{\begin{tabular}[c]{@{}c@{}}Fault freq. \\ {[}Hz{]}\end{tabular}} & \textbf{\begin{tabular}[c]{@{}c@{}}Speed \\ {[}rpm{]}\end{tabular}} & \textbf{\begin{tabular}[c]{@{}c@{}}d\\ {[}mm{]}\end{tabular}} & \textbf{\begin{tabular}[c]{@{}c@{}}D\\ {[}mm{]}\end{tabular}} & \textbf{\begin{tabular}[c]{@{}c@{}}B\\ {[}mm{]}\end{tabular}} \\ \hline
SKF 1205 EKTN9   & 91.5                                                                     & 1000                                                                & 25                                                            & 52                                                            & 15                                                            \\
SKF 6204-2RSH    & 5.4                                                                      & 105-110                                                             & 20                                                            & 47                                                            & 14                                                            \\ \hline
\end{tabular}
\end{table}

\subsubsection{Gaussian noise}
\label{subsub:gaussian_noise}
\begin{figure}[h!]
    \centering
    \includegraphics[scale=0.3]{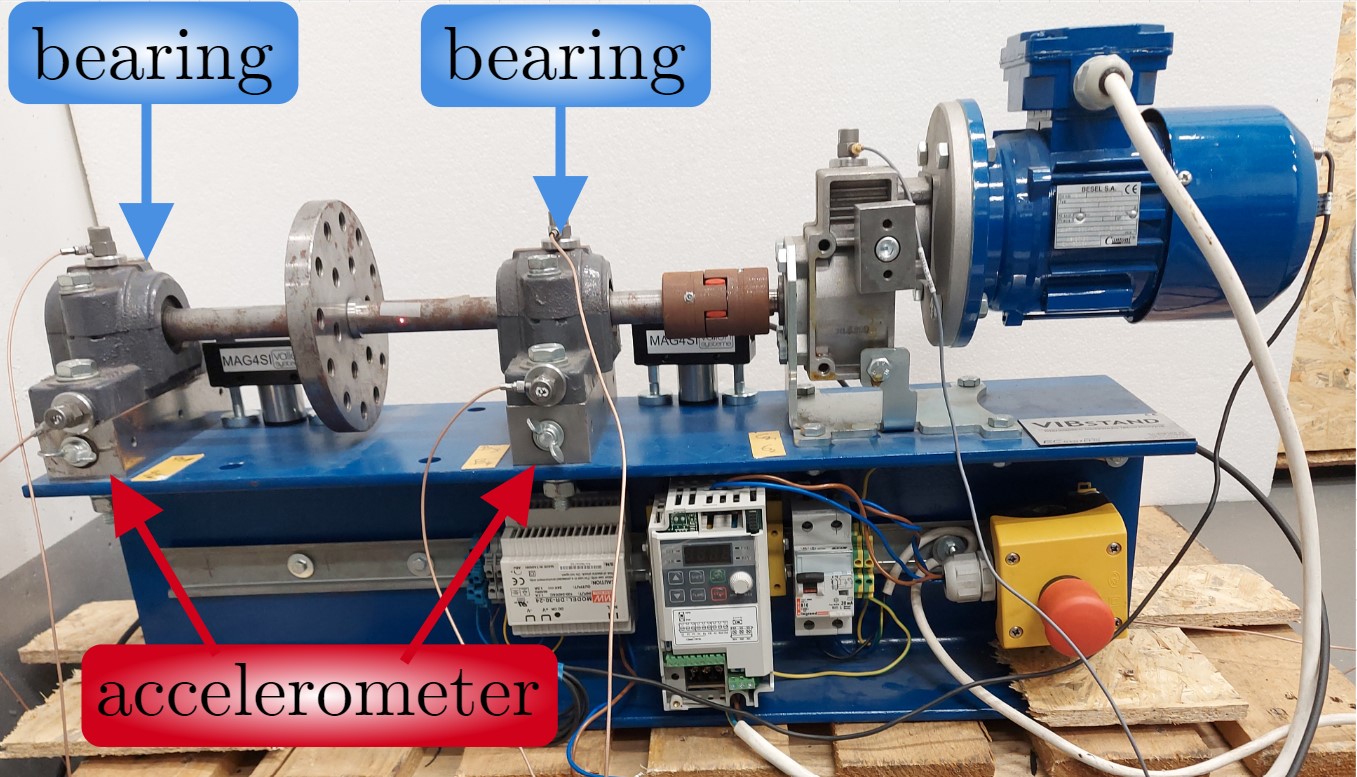}
    \caption{Test rig used in experiments.}
    \label{fig:test_rig2}
\end{figure}

\begin{figure}[h!]
    \centering
    \begin{subfigure}[b]{0.49\linewidth}
         \centering
         \includegraphics[width=\linewidth]{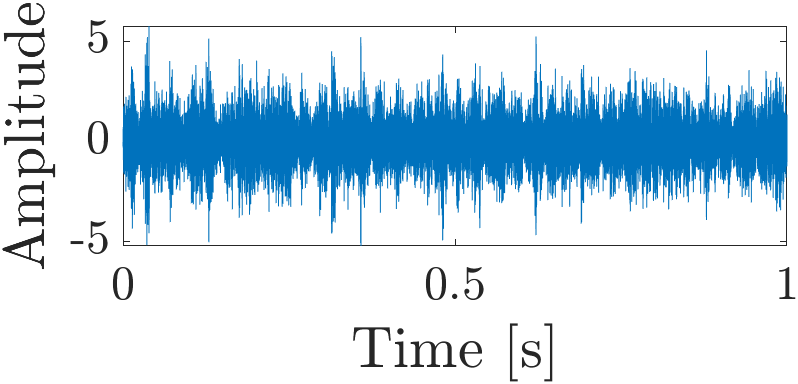}
         \caption{Gaussian noise}
     \end{subfigure}
     \hfill
     \begin{subfigure}[b]{0.49\linewidth}
         \centering
         \includegraphics[width=\linewidth]{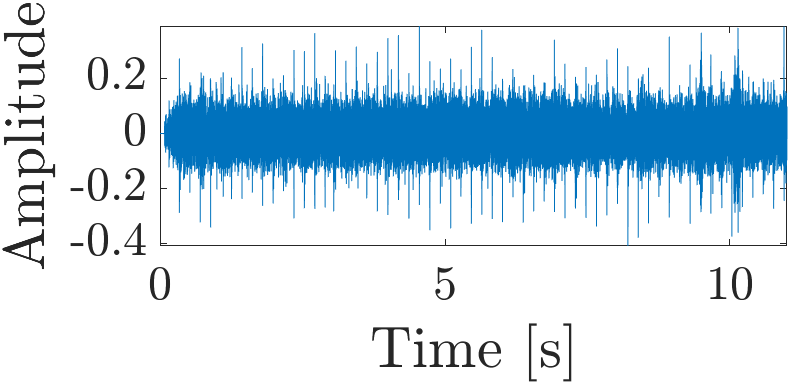}
         \caption{non-Gaussian noise}
     \end{subfigure}
     \vskip\baselineskip
     \begin{subfigure}[b]{0.49\linewidth}
         \centering
         \includegraphics[width=\linewidth]{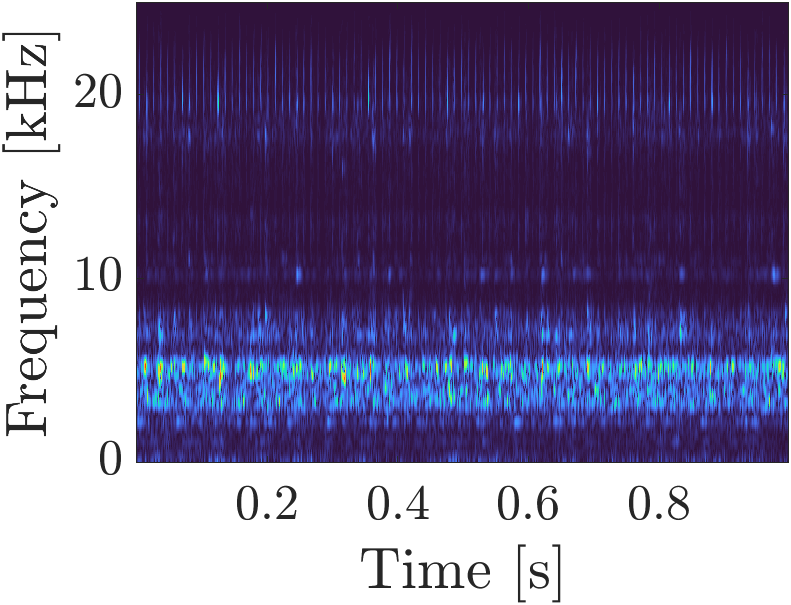}
         \caption{Gaussian noise}
     \end{subfigure}
     \hfill
     \begin{subfigure}[b]{0.49\linewidth}
         \centering
         \includegraphics[width=\linewidth]{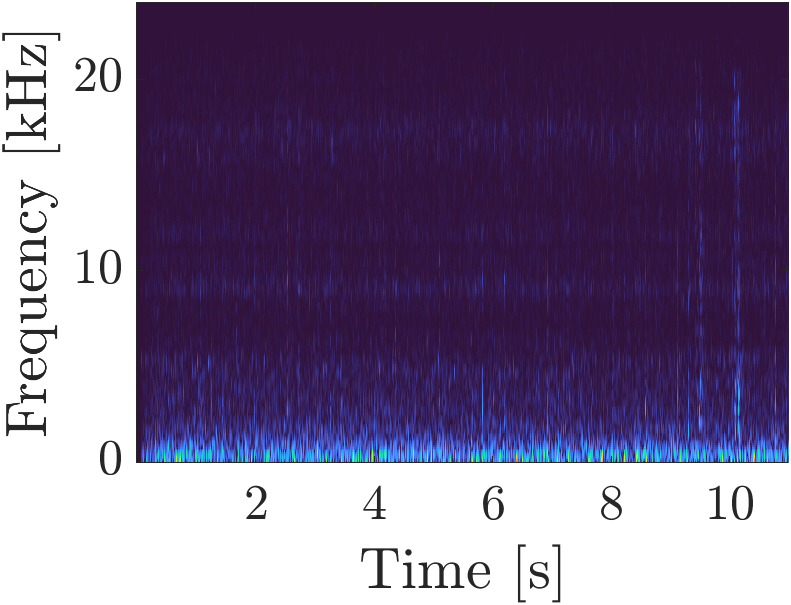}
         \caption{non-Gaussian noise}
     \end{subfigure}

    \caption{Recorded real vibration signal with Gaussian noise (a), its spectrogram (c) and real acoustic signal with non-Gaussian noise (b), its spectrogram (d).}
    \label{fig:soi_real}
\end{figure}

A test rig (presented in Figure \ref{fig:test_rig2}) was used to capture a vibration signal. The test rig contains an electric motor, one-stage gearbox coupling, and two bearings. One of them is intentionally damaged. The KISTLER Model 8702B500 accelerators were stacked horizontally (orthogonally to the axis of rotation) and vertically to each bearing. The signal of the faulty bearing was captured with the sampling rate of 50 kHz. The recorded signal and its spectrogram is presented in Figure \ref{fig:soi_real} (a) and Figure \ref{fig:soi_real} (c), accordingly. Cyclic impulses are not clearly visible in raw vibrations. However, the spectrogram of the signal allows us to notice an informative band around 20 kHz. 

  
\begin{figure}[h!]
    \centering
    \includegraphics[scale=0.3]{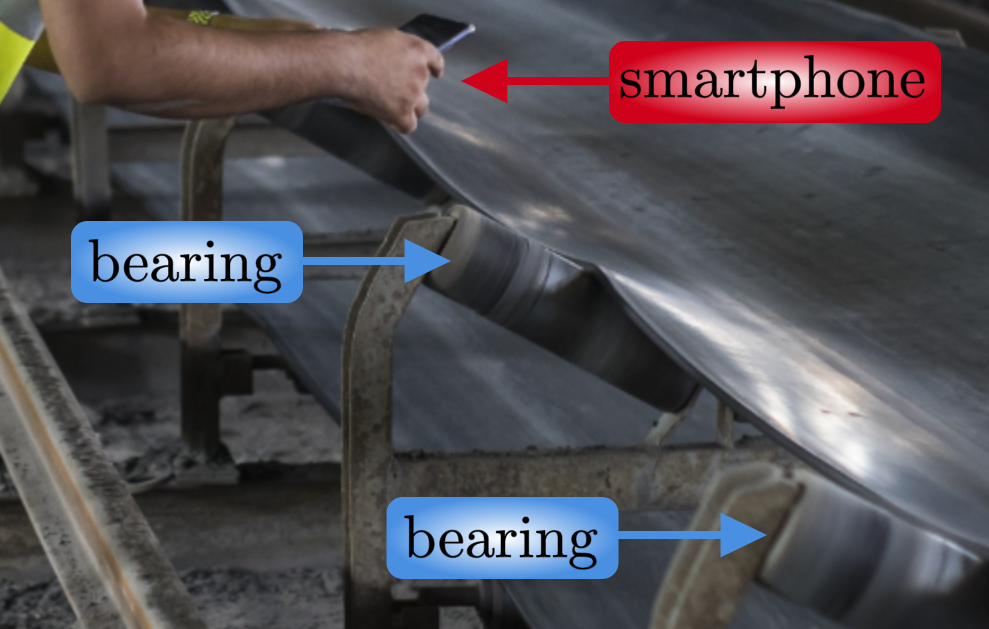}
    \caption{Sound measurement.}
    \label{fig:field_measurement}
\end{figure}

\subsubsection{Non-Gaussian noise}
\label{subsub:nongaussian_noise}
The second real signal is related to sound measurement. Idlers are a crucial part of a belt conveyor. The condition monitoring of rolling element bearings installed on idlers, due to their rotation, is possible by analysis of infrared thermographic or sound rather than by vibration processing. In this study, we used the audio recordings made with a smartphone, which is presented in Figure \ref{fig:field_measurement}. The measurement contains an 11-second excerpt of the acoustic signal sampled with 48 kHz.

 The acoustic signal is depicted in Figure \ref{fig:soi_real} (b) and its spectrogram in Figure \ref{fig:soi_real} (d). Although cyclic impulses are visible in the raw signal, there is some noise. The spectrogram of the signal showed several wideband impulses that are considered as random disturbances. They make band selection quite complicated. Rough visual inspection suggests the location of an informative band around 2-5kHz. The non-Gaussian impulses are mostly visible in the range from about the 9th to the 10th second, encompassing the whole frequency bands.


\section{\textcolor{black}{Results}}
\label{sub:results}

\textcolor{black}{In this section, the proposed method is compared with the selected state-of-the-art methods, and the results are presented for both simulated and real signals. We used two types of real signals: the vibration signal with Gaussian noise, and the acoustic signals with non-Gaussian noise.
}

\subsection{Simulated signals}
\label{subsub:simulated_results}

\begin{figure}[h!]
    \centering
    \includegraphics[scale=0.32]{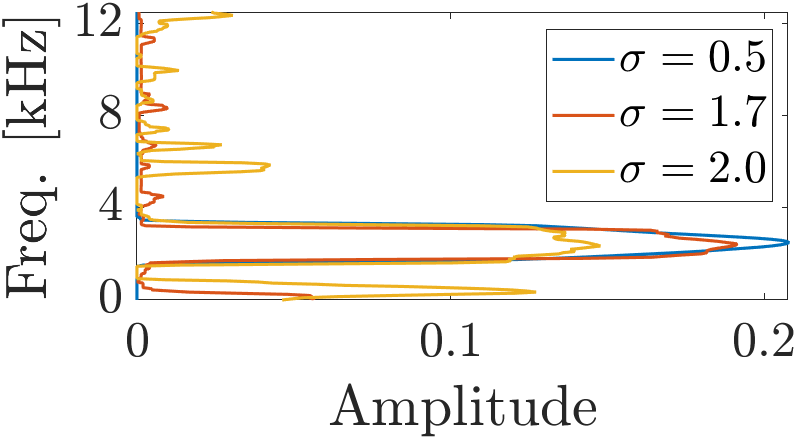}
    \caption{Averaged filters obtained by SS-ONMF on simulated signals with Gaussian noise.}
    \label{fig:gauss_filters}
\end{figure}

\begin{figure}
     \centering
     \begin{subfigure}[b]{0.24\textwidth}
         \centering
         \includegraphics[width=\textwidth]{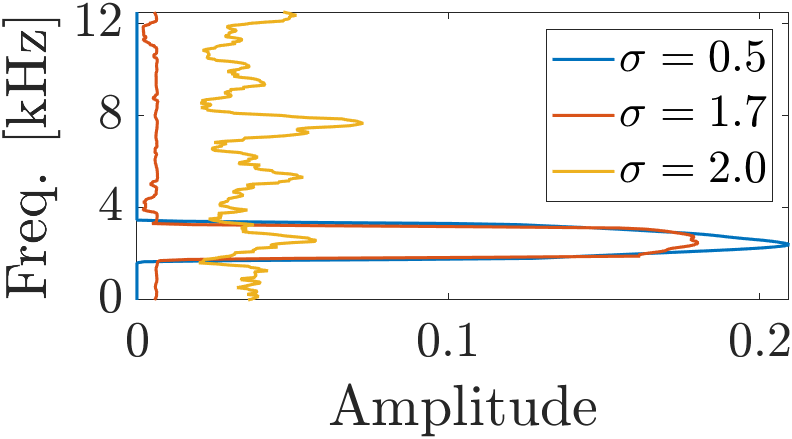}
         \caption{}
         \label{fig:filters_5}
     \end{subfigure}
     \hfill
     \begin{subfigure}[b]{0.24\textwidth}
         \centering
         \includegraphics[width=\textwidth]{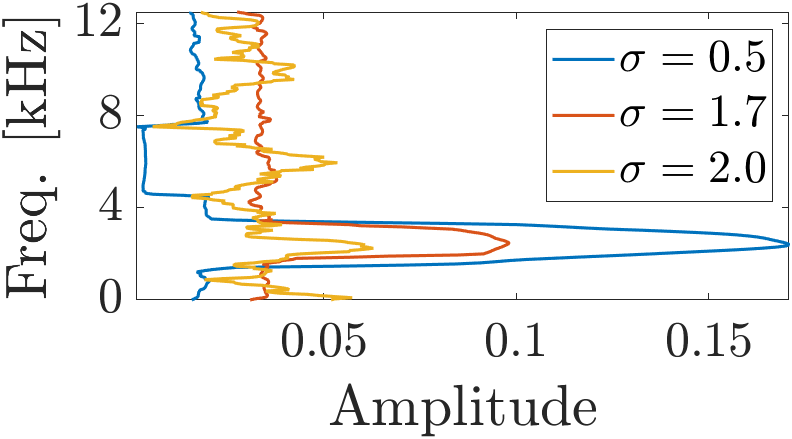}
         \caption{}
         \label{fig:filters_10}
     \end{subfigure}
     
     \caption{Averaged filters obtained by SS-ONMF on simulated signals with non-Gaussian noise: (a) max. amplitude of non-Gaussian impulses equal to 5 and (b) max. amplitude of non-Gaussian impulses equal to 15.}
    \label{fig:non_gauss_filters}
\end{figure}

The average frequency characteristics of the obtained filters with the best rank for simulated signals with Gaussian and non-Gaussian noises are depicted in Figures \ref{fig:gauss_filters} and \ref{fig:non_gauss_filters}, respectively. Figure \ref{fig:gauss_filters} shows that the proposed approach finds the right frequency band. Obviously, an increase in the noise power diminishes the quality of the estimated filters. However, even if the SOI is completely hidden in the Gaussian noise, the SS-ONMF is still able to find the OFB. Some spectral leakage could be further eliminated by thresholding.

The non-Gaussian case presented in Figure \ref{fig:non_gauss_filters} is more complicated. In the case of small non-cyclic impulses, the SS-ONMF works similarly to the Gaussian case. Unfortunately, large non-cyclic impulses, together with the Gaussian noise that masks the SOI, make the proposed methodology insufficient in the worst-considered scenario.

\subsection{Real signals}
\label{subsub:real_results}
 For the real signals, we selected a single filter whose the kurtosis/ENVSI value of the filtered signal is equal to the median of the obtained results with the best rank for both signals with Gaussian  \textcolor{black}{(vibration signal)} and non-Gaussian noise \textcolor{black}{(acoustic signal}). The frequency characteristics of the filters produced by the compared methods and the corresponding filtered real signals with Gaussian and non-Gaussian noise are presented in Figures \ref{fig:best_gauss} and \ref{fig:best_non_gauss}, respectively.

 \subsubsection{\textcolor{black}{Vibration signal}}
\begin{figure}[]
     \centering
     
     \begin{subfigure}[b]{\linewidth}
        \caption{SS-ONMF, Kurtosis: 28.133}
        \centering
        \includegraphics[scale=0.3]{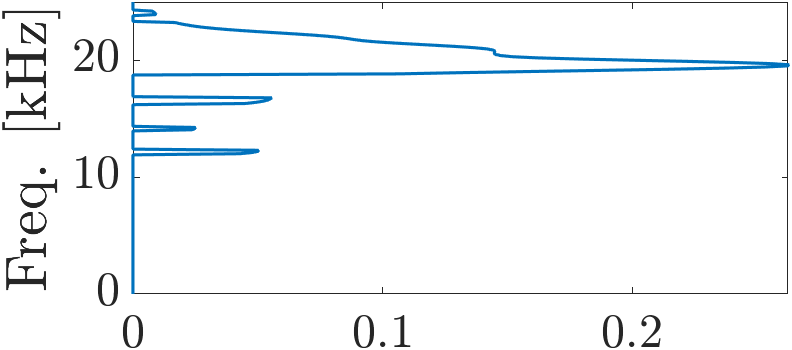}%
        \hfill
        \includegraphics[scale=0.3]{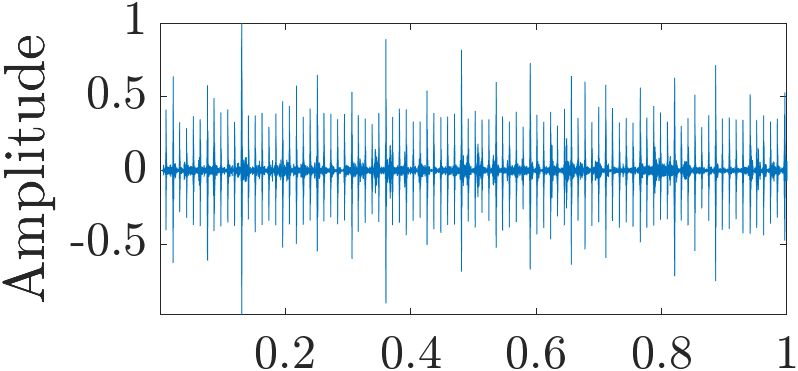}
    \end{subfigure}
    
    \begin{subfigure}[b]{\linewidth}
        \caption{ONMFS, Kurtosis: 4.074}
        \centering
        \includegraphics[scale=0.3]{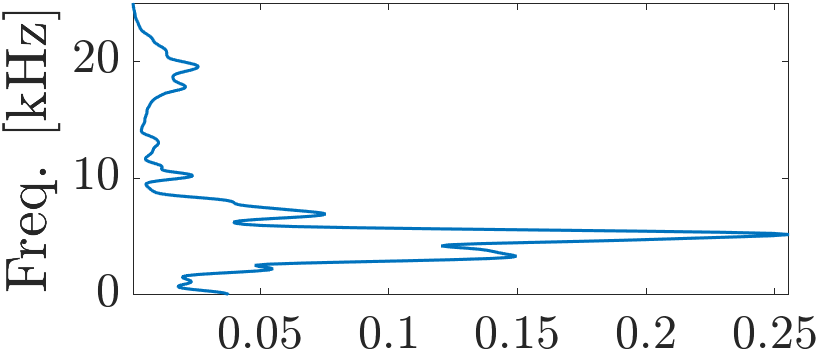}%
        \hfill
        \includegraphics[scale=0.3]{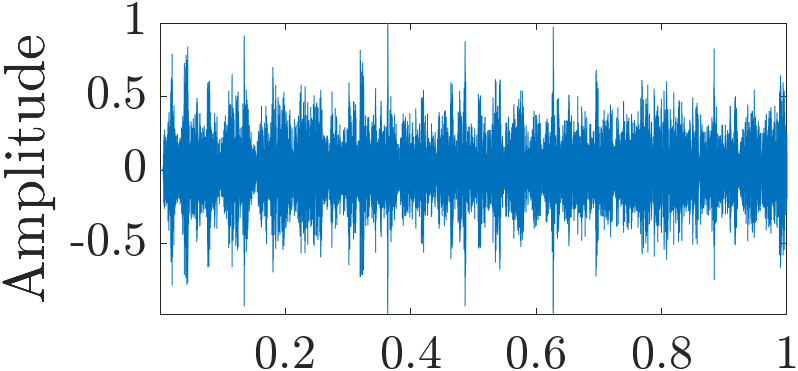}
    \end{subfigure}
    
    \begin{subfigure}[b]{\linewidth}
        \caption{NMF-MU, Kurtosis: 11.657}
        \centering
        \includegraphics[scale=0.3]{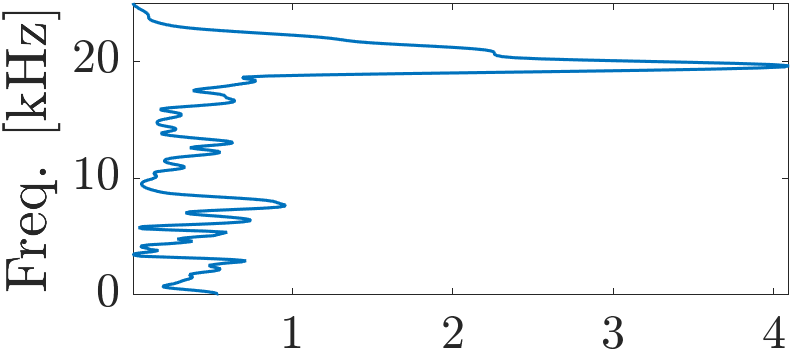}%
        \hfill
        \includegraphics[scale=0.3]{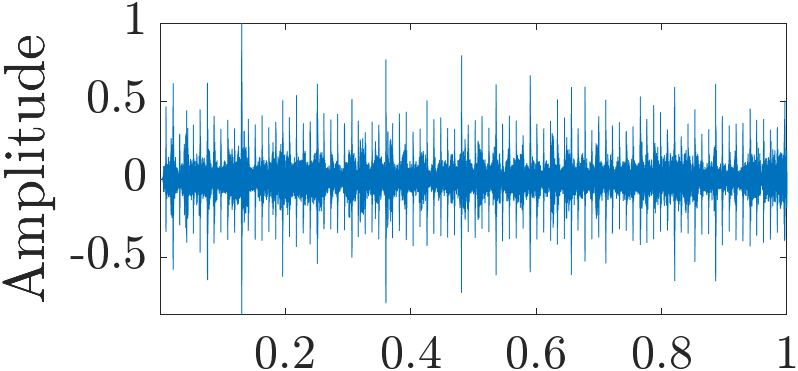}
    \end{subfigure}

    \begin{subfigure}[b]{\linewidth}
        \caption{CVB, Kurtosis: 5.039}
        \centering
        \includegraphics[scale=0.3]{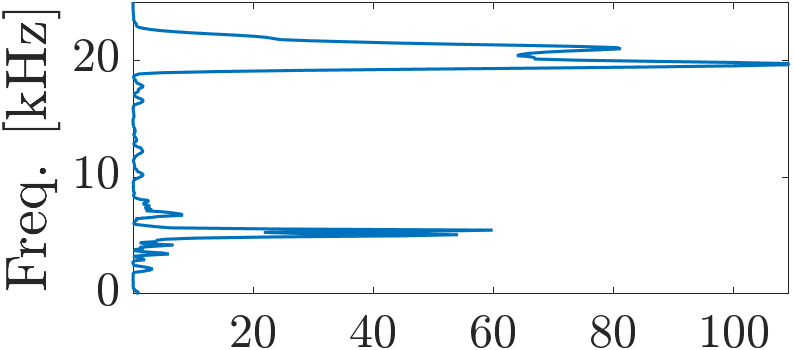}%
        \hfill
        \includegraphics[scale=0.3]{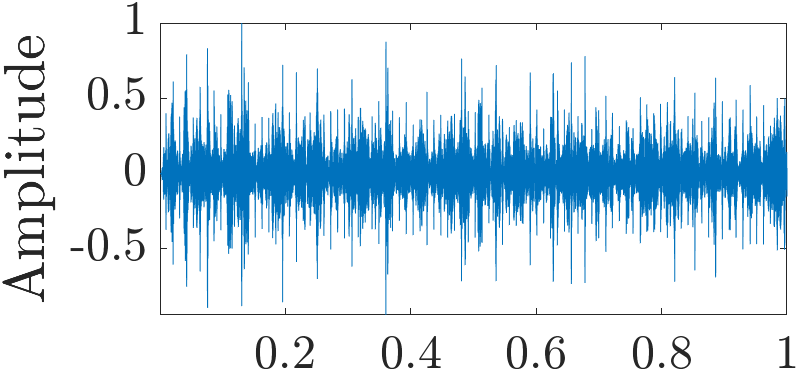}
    \end{subfigure}

    \begin{subfigure}[b]{\linewidth}
        \caption{SK, Kurtosis: 5.479}
        \centering
        \includegraphics[scale=0.3]{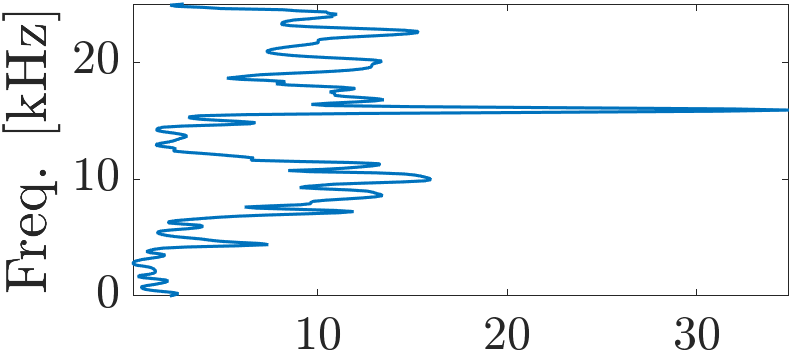}%
        \hfill
        \includegraphics[scale=0.3]{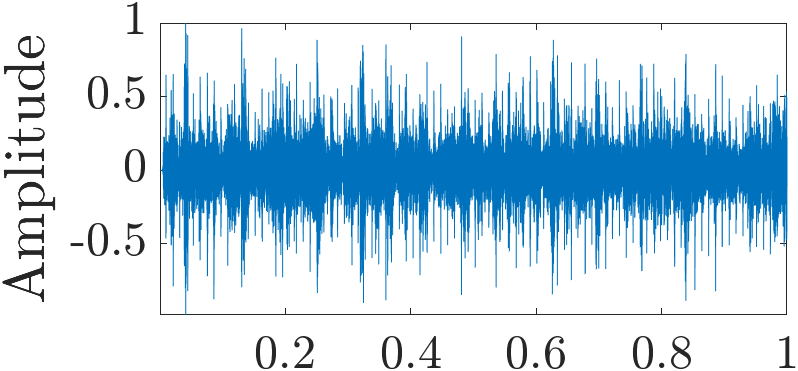}
    \end{subfigure}

    \begin{subfigure}[b]{\linewidth}
        \caption{infogram, Kurtosis: 5.340}
        \centering
        \includegraphics[scale=0.3]{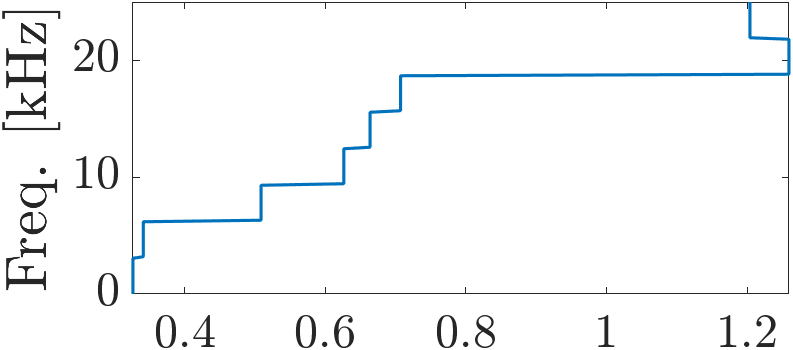}%
        \hfill
        \includegraphics[scale=0.3]{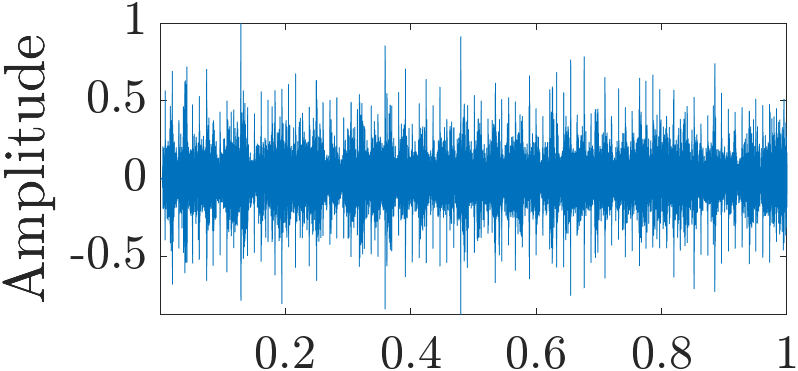}
    \end{subfigure}

    \begin{subfigure}[b]{\linewidth}
        \caption{sparsogram, Kurtosis: 4.030}
        \centering
        \includegraphics[scale=0.3]{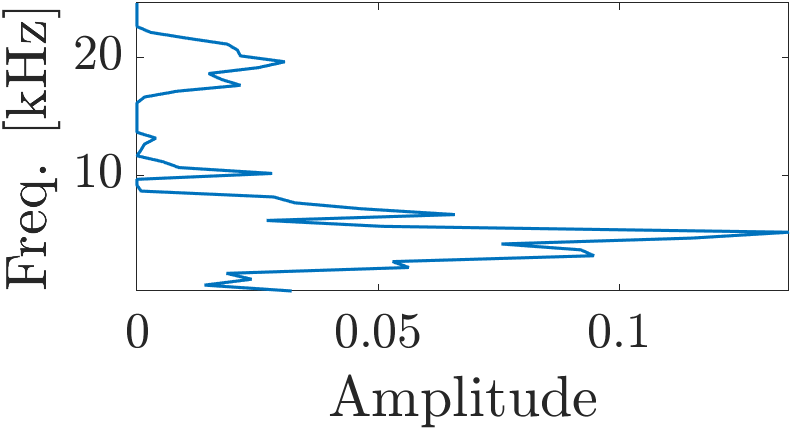}%
        \hfill
        \includegraphics[scale=0.3]{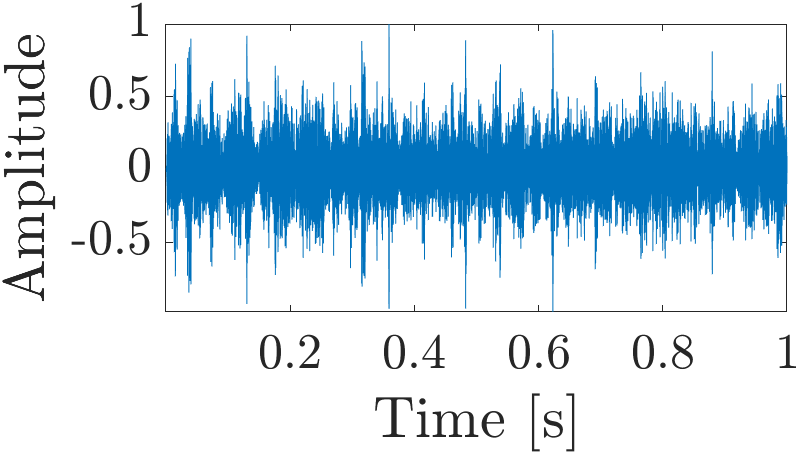}
    \end{subfigure}
    
    \caption{Exemplary filters and real Gaussian noisy signals filtered with analyzed methods.}
    \label{fig:best_gauss}
\end{figure}

 Figure \ref{fig:best_gauss} shows that SS-ONMF and NMF-MU yield similar OFB, however, SS-ONMF provides the smallest spectral leakage. Consequently, the filtered signals with SS-ONMF have the highest value of the kurtosis. It results both from the orthogonality constraints as well as a better stability of SS-ONMF with reference to ONMFS. It is worthy to notice that the filter characteristic obtained with the infogram looks differently than the others. It is related to the number of decomposition levels (we used infogram values for the whole frequency range to build the filter). The frequency resolution for all the methods depends on the resolution of the spectrogram, but in the case of the infogram - it depends on the selected number of filters. \textcolor{black}{Sparsogram properly identifies the optimal frequency band around 20 kHz, but it also identifies the non-informative frequency band with a higher amplitude, which results in a lower-quality filter.}
 
\begin{figure}[]
     \centering
     
     \begin{subfigure}[b]{\linewidth}
        \caption{Envelope Spectrum - SS-ONMF, ENVSI: 0.470}
        \centering
        \includegraphics[scale=0.32]{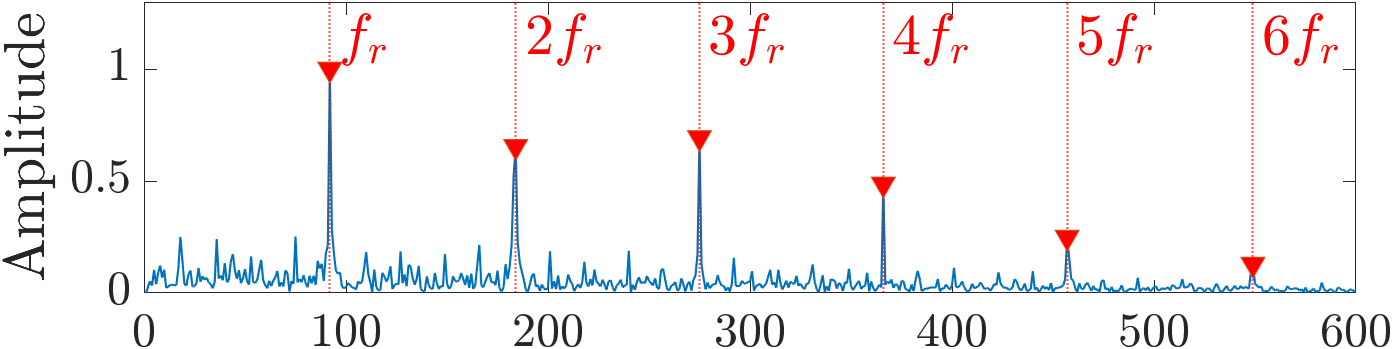}
    \end{subfigure}
    
    \begin{subfigure}[b]{\linewidth}
        \caption{Envelope Spectrum - ONMFS, ENVSI: 0.164}
        \centering
        \includegraphics[scale=0.32]{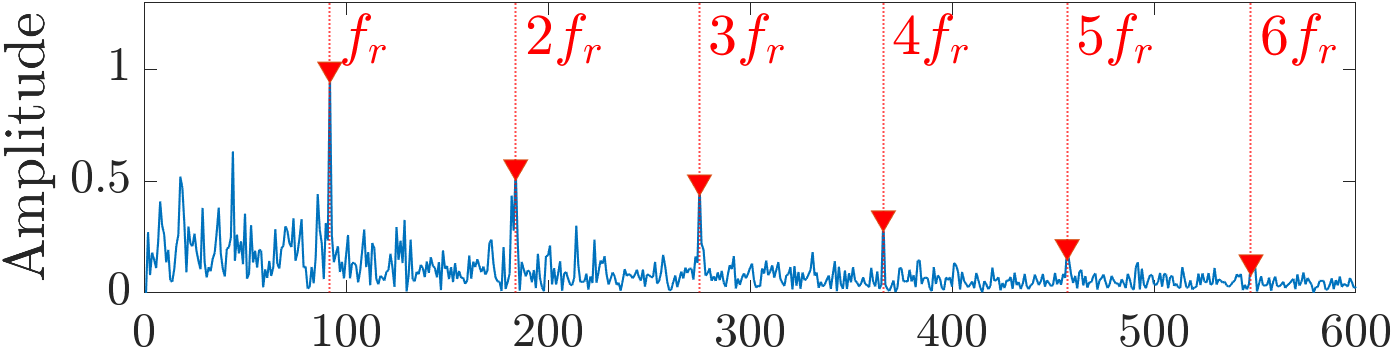}
    \end{subfigure}
    
    \begin{subfigure}[b]{\linewidth}
        \caption{Envelope Spectrum - NMF-MU, ENVSI: 0.448}
        \centering
        \includegraphics[scale=0.32]{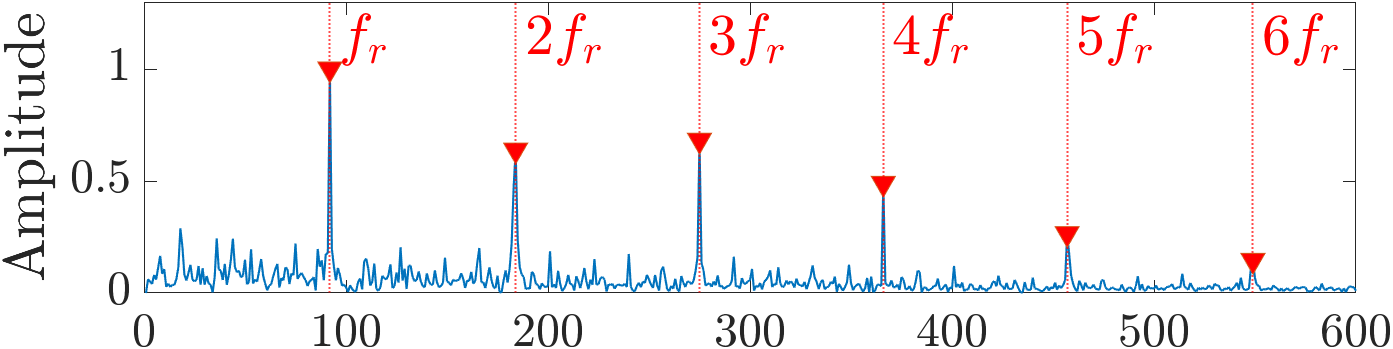}
    \end{subfigure}

    \begin{subfigure}[b]{\linewidth}
        \caption{Envelope Spectrum - CVB, ENVSI: 0.491}
        \centering
        \includegraphics[scale=0.32]{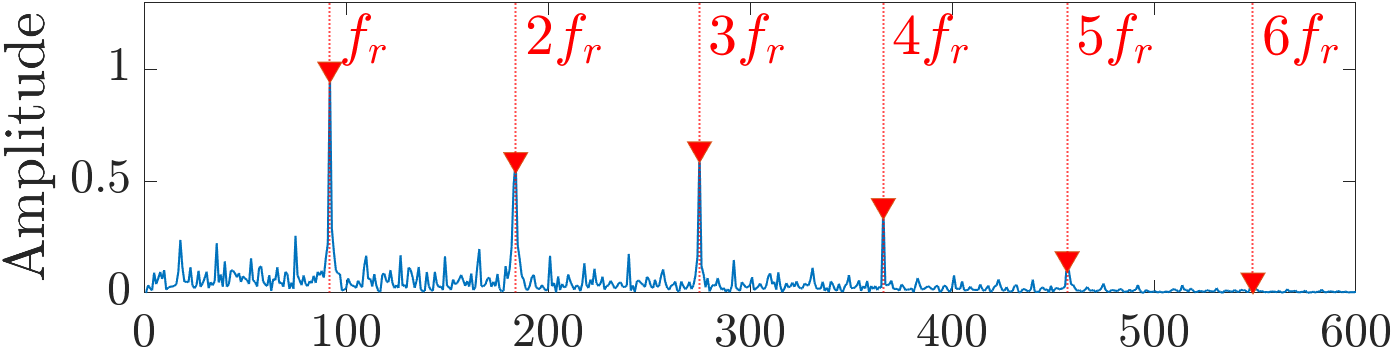}
    \end{subfigure}

    \begin{subfigure}[b]{\linewidth}
        \caption{Envelope Spectrum - SK, ENVSI: 0.164}
        \centering
        \includegraphics[scale=0.32]{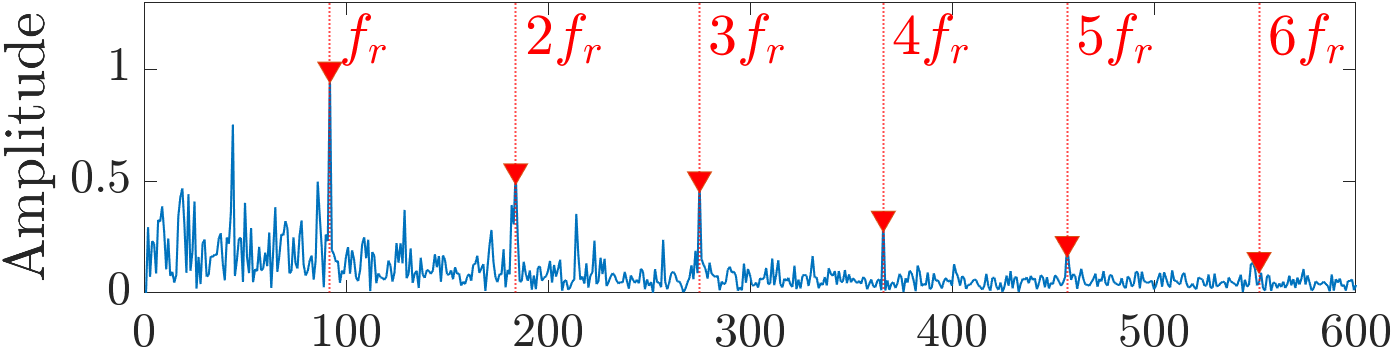}
    \end{subfigure}

    \begin{subfigure}[b]{\linewidth}
        \caption{Envelope Spectrum - infogram, ENVSI: 0.274}
        \centering
        \includegraphics[scale=0.32]{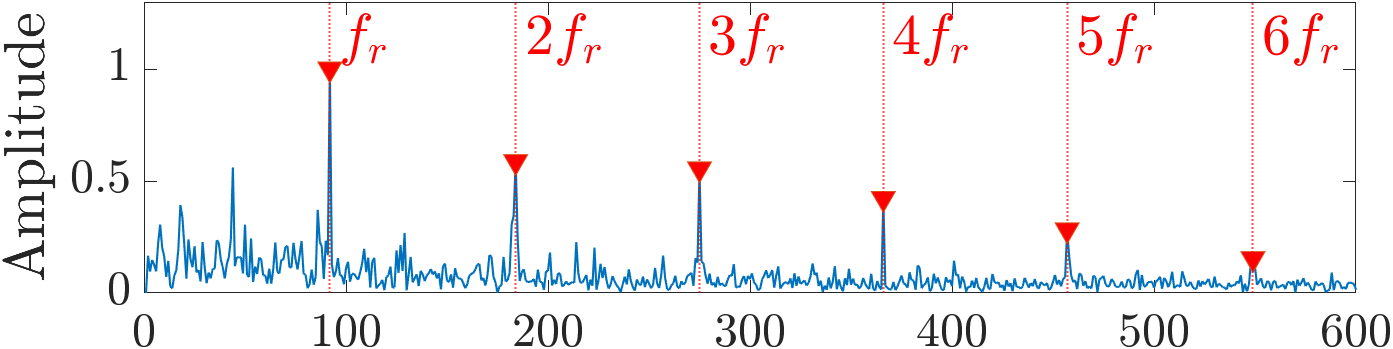}
    \end{subfigure}

    \begin{subfigure}[b]{\linewidth}
        \caption{Envelope Spectrum - sparsogram, ENVSI: 0.148}
        \centering
        \includegraphics[scale=0.32]{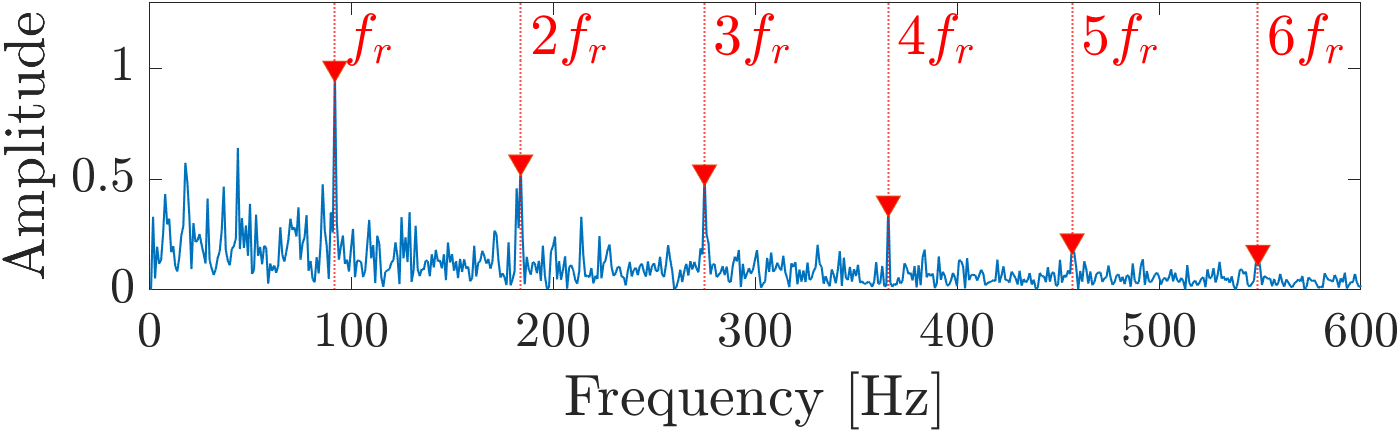}
    \end{subfigure}
    
    \caption{Envelope spectra of real signal with Gaussian noise filtered with analyzed methods.}
    \label{fig:envelope_spectra_gauss}     
\end{figure}

 The final step in condition monitoring is analysis of the envelope spectrum and the identification of the components that correspond to the so-called fault frequency and its harmonics. These frequencies are marked with the arrows in Figure \ref{fig:envelope_spectra_gauss}. For Gaussian noise, all the methods allow to detect damage. Differences in envelope spectrum are minor and are related to the background noise and amplitudes of higher harmonics. The ENVSI calculated for the envelope spectra in Figure \ref{fig:envelope_spectra_gauss} is the highest for CVB, but is only slightly higher than for SS-ONMF. However, the filtered signal obtained with SS-ONMF in the time domain is significantly better than with CVB. The filter characteristic is also better (more selective) for SS-ONMF than for CVB.

\subsubsection{\textcolor{black}{Acoustic signal}}
\begin{figure}[]
     \centering
     
     \begin{subfigure}[b]{\linewidth}
        \caption{SS-ONMF}
        \centering
        \includegraphics[scale=0.3]{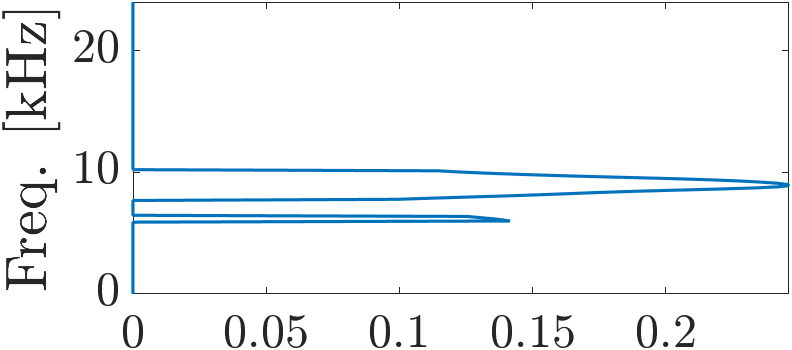}
        \hfill
        \includegraphics[scale=0.3]{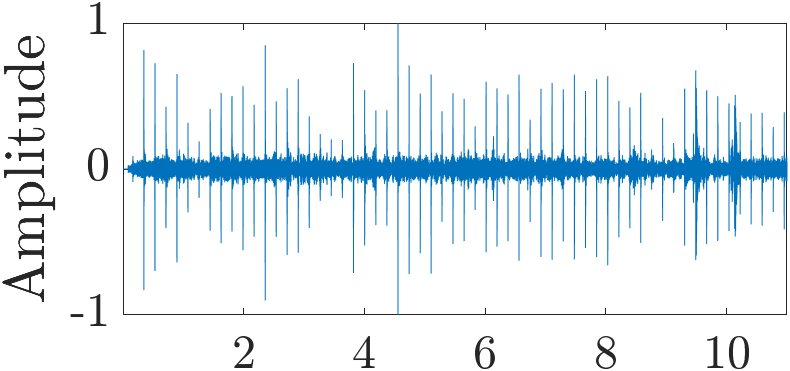}
        
    \end{subfigure}
    
    \begin{subfigure}[b]{\linewidth}
        \caption{ONMFS}
        \centering
        \includegraphics[scale=0.3]{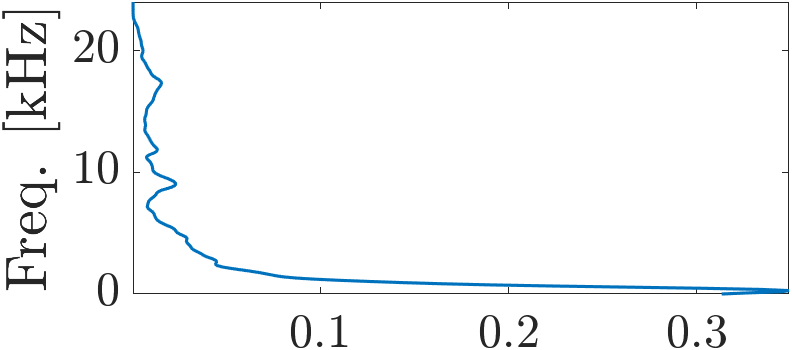}
        \hfill
        \includegraphics[scale=0.3]{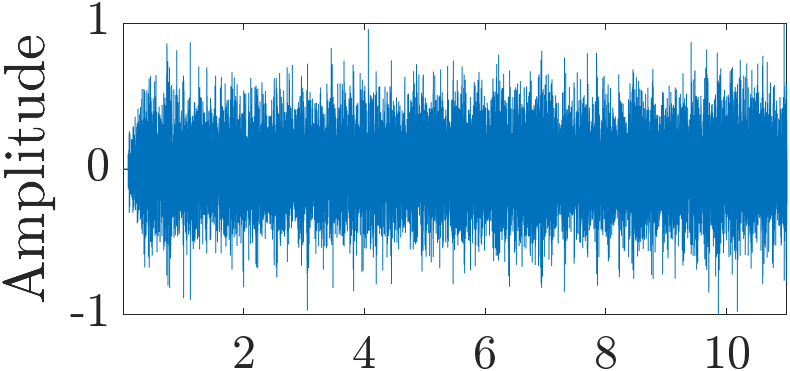}
    \end{subfigure}
    
    \begin{subfigure}[b]{\linewidth}
        \caption{NMF-MU}
        \centering
        \includegraphics[scale=0.3]{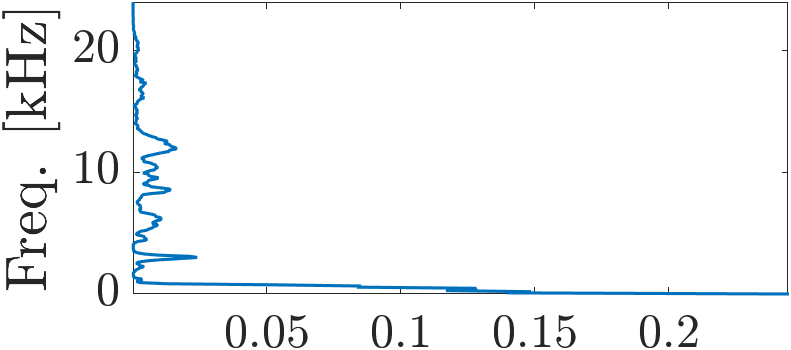}
        \hfill
        \includegraphics[scale=0.3]{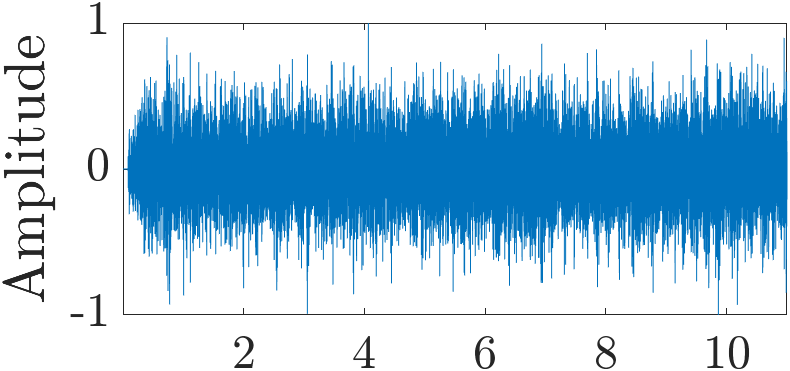}
    \end{subfigure}

    \begin{subfigure}[b]{\linewidth}
        \caption{CVB}
        \centering
        \includegraphics[scale=0.3]{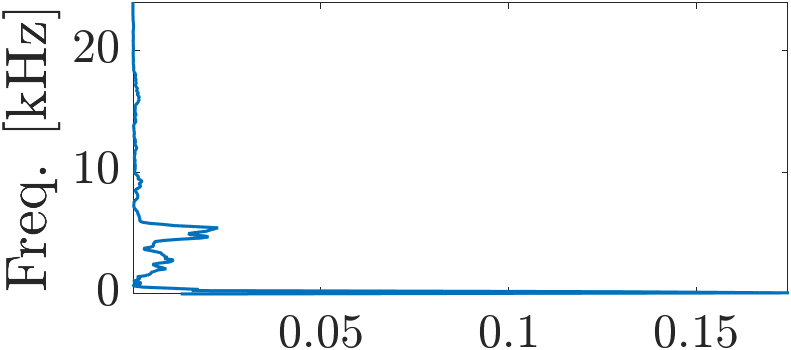}
        \hfill
        \includegraphics[scale=0.3]{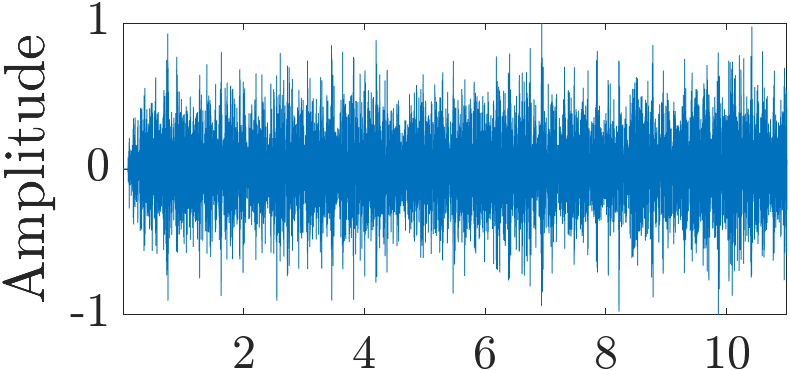}
    \end{subfigure}

    \begin{subfigure}[b]{\linewidth}
        \caption{SK}
        \centering
        \includegraphics[scale=0.3]{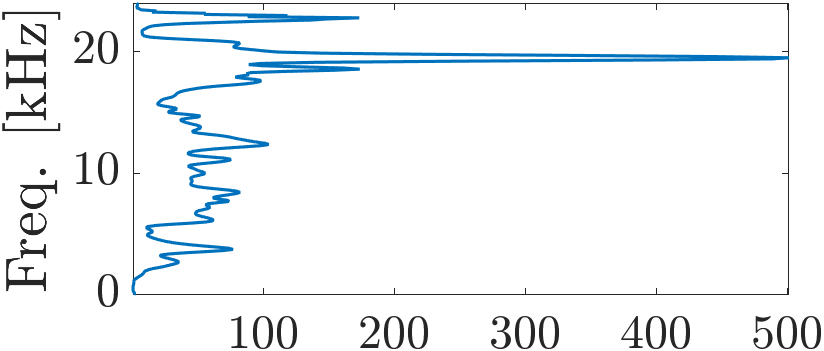}
        \hfill
        \includegraphics[scale=0.3]{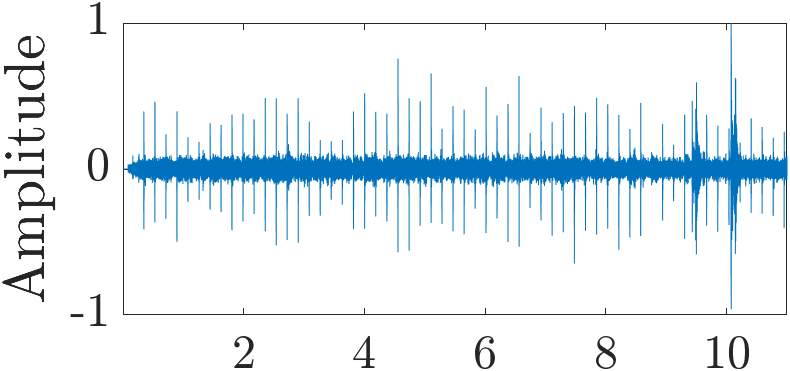}
    \end{subfigure}

    \begin{subfigure}[b]{\linewidth}
        \caption{infogram}
        \centering
        \includegraphics[scale=0.3]{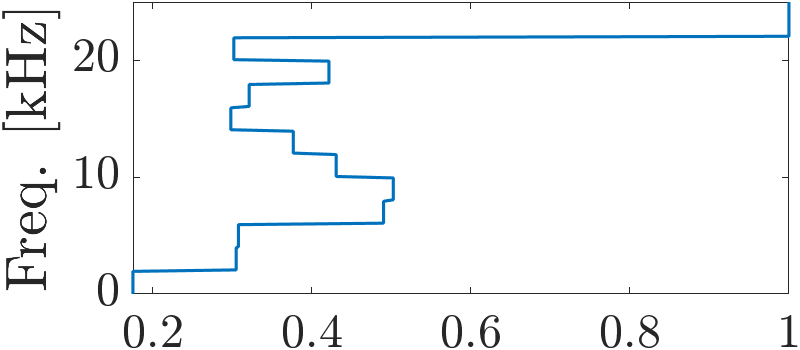}
        \hfill
        \includegraphics[scale=0.3]{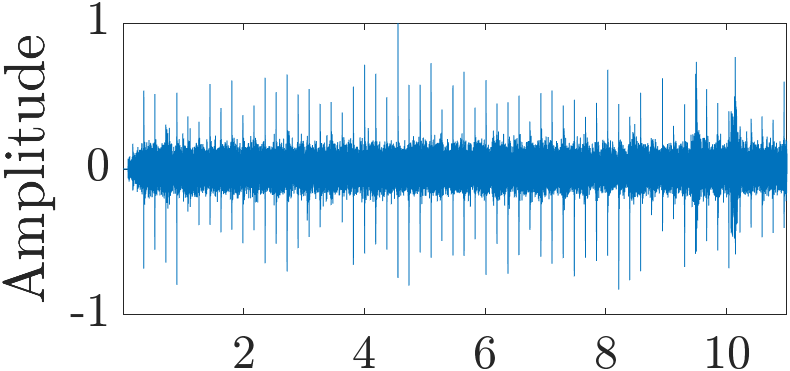}
    \end{subfigure}

    \begin{subfigure}[b]{\linewidth}
        \caption{sparsogram}
        \centering
        \includegraphics[scale=0.3]{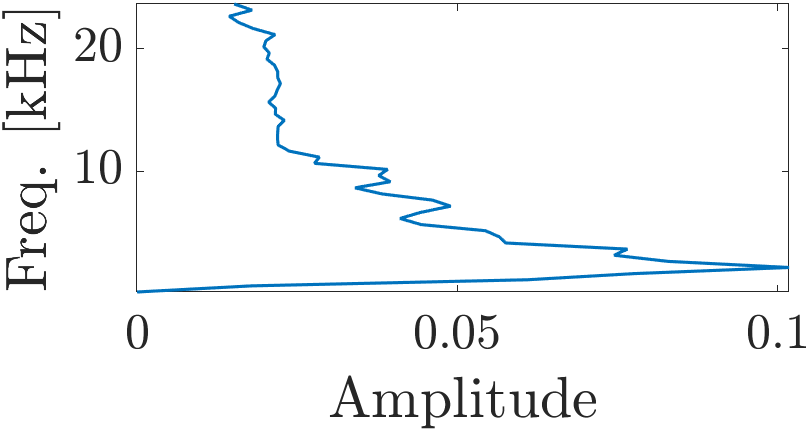}
        \hfill
        \includegraphics[scale=0.3]{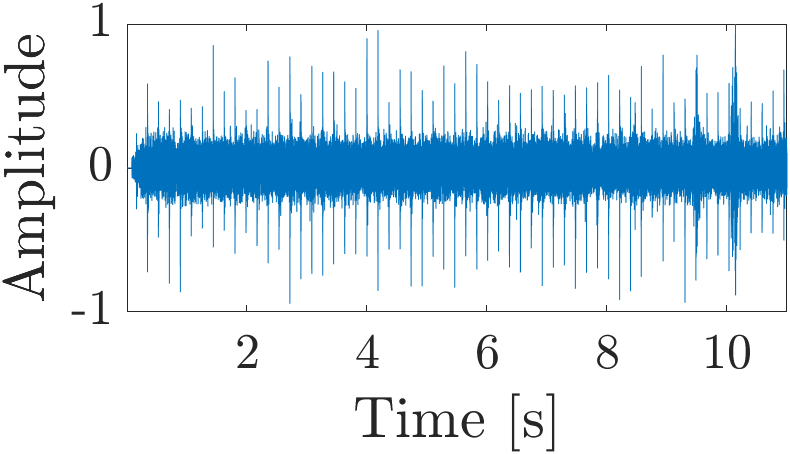}
    \end{subfigure}
    
    \caption{Exemplary filters and real non-Gaussian noisy signals filtered with analyzed methods.}
    \label{fig:best_non_gauss}     
\end{figure}

The performance of the considered methods for filtering the acoustic signal with non-Gaussian disturbances can be observed in Figure \ref{fig:best_non_gauss}. In this case, we cannot use the kurtosis to evaluate the filtered signal because it is not suitable for non-Gaussian disturbances. In consequence, we evaluated the envelope spectra of the filtered signals using the ENVSI measure. 
As can be noted, SS-ONMF provides the most selective filter 
which directly points at an informative frequency band. The other methods yield more complicated (multi-modal) filters, and the filtered signals obtained with ONMFS, CVB and NMF-MU are much more noisy. \textcolor{black}{In the case of sparsogram, the filtered signal is impulsive and cyclic but strongly perturbed by the background noise leaking from the non-informative band, which results into a low ENVSI value.} The most similar results are obtained by SK and infogram. With respect to the analysis of the envelope spectra, SS-ONMF gives the highest ENVSI from all the compared methods. For the non-Gaussian case, SS-ONMF is the only method which provides very good results in both the time and frequency domains.

This observation confirms our assumption that the orthogonality constraints imposed to NMF are crucial in our application to extract the right filter.

\begin{figure}[]
     \centering
     
     \begin{subfigure}[b]{\linewidth}
        \caption{Envelope Spectrum - SS-ONMF, ENVSI: 0.191}
        \centering
        \includegraphics[scale=0.32]{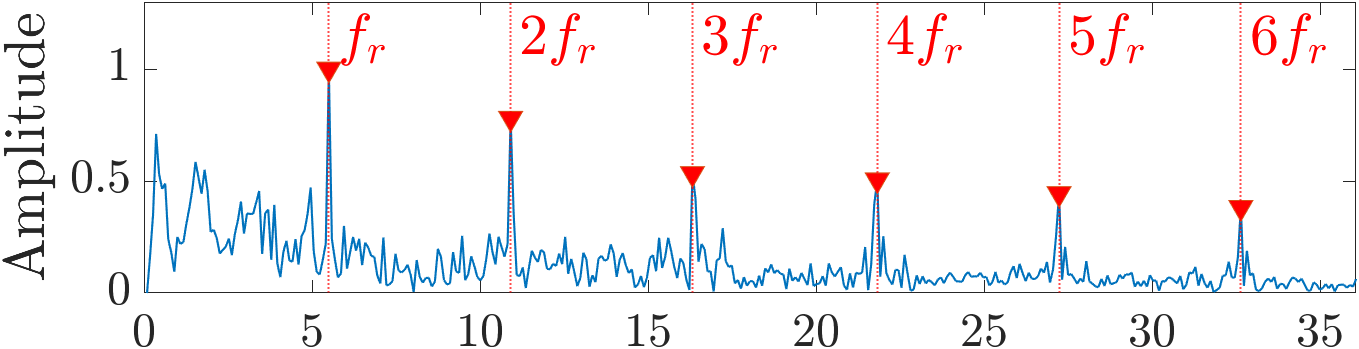}
    \end{subfigure}
    
    \begin{subfigure}[b]{\linewidth}
        \caption{Envelope Spectrum - ONMFS, ENVSI: 0.072}
        \centering
        \includegraphics[scale=0.32]{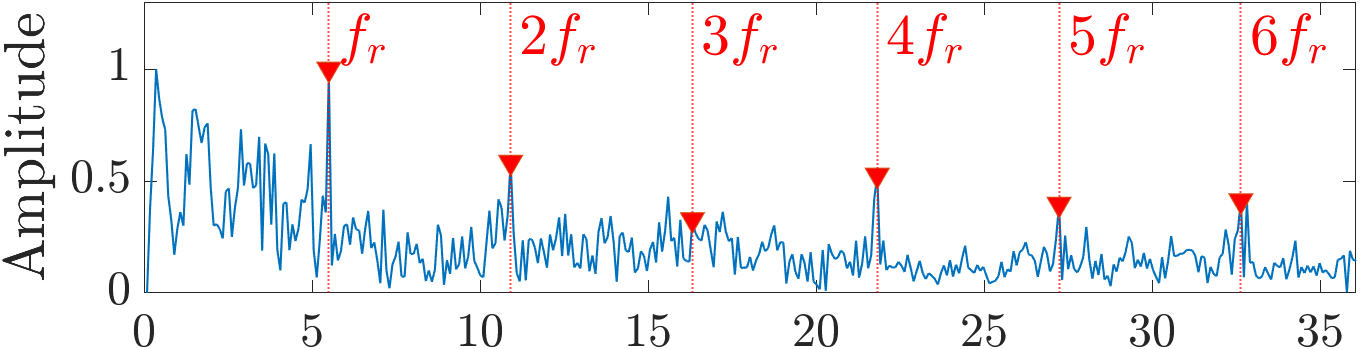}
    \end{subfigure}
    
    \begin{subfigure}[b]{\linewidth}
        \caption{Envelope Spectrum - NMF-MU, ENVSI: 0.138}
        \centering
        \includegraphics[scale=0.32]{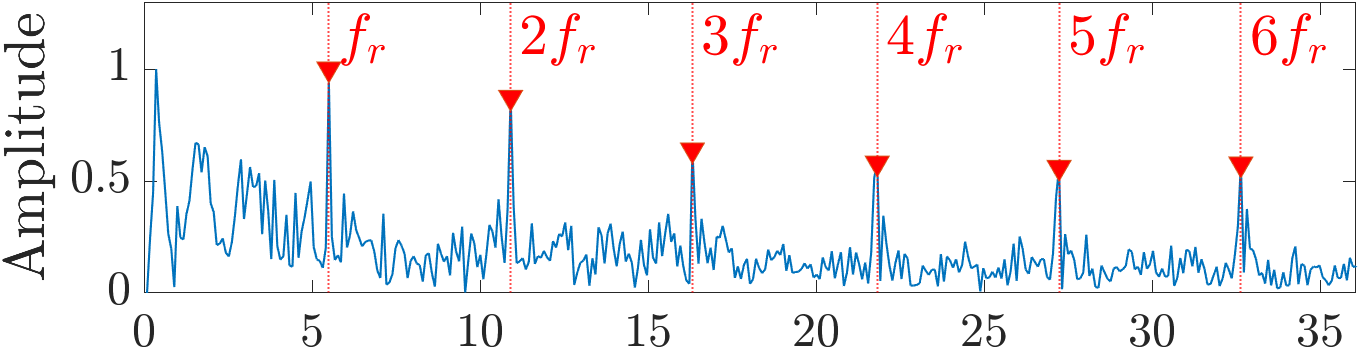}
    \end{subfigure}

    \begin{subfigure}[b]{\linewidth}
        \caption{Envelope Spectrum - CVB, ENVSI: 0.079}
        \centering
        \includegraphics[scale=0.32]{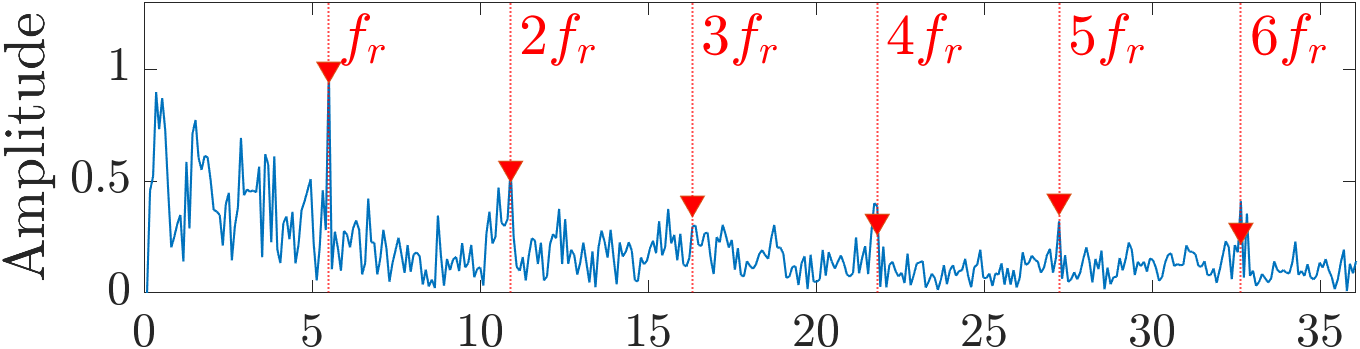}
    \end{subfigure}

    \begin{subfigure}[b]{\linewidth}
        \caption{Envelope Spectrum - SK, ENVSI: 0.089}
        \centering
        \includegraphics[scale=0.32]{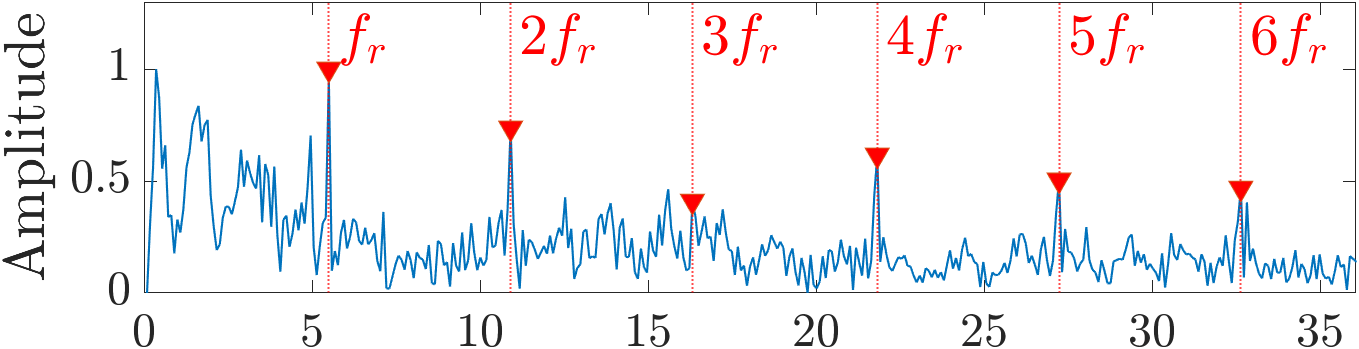}
    \end{subfigure}

    \begin{subfigure}[b]{\linewidth}
        \caption{Envelope Spectrum - infogram, ENVSI: 0.093}
        \centering
        \includegraphics[scale=0.32]{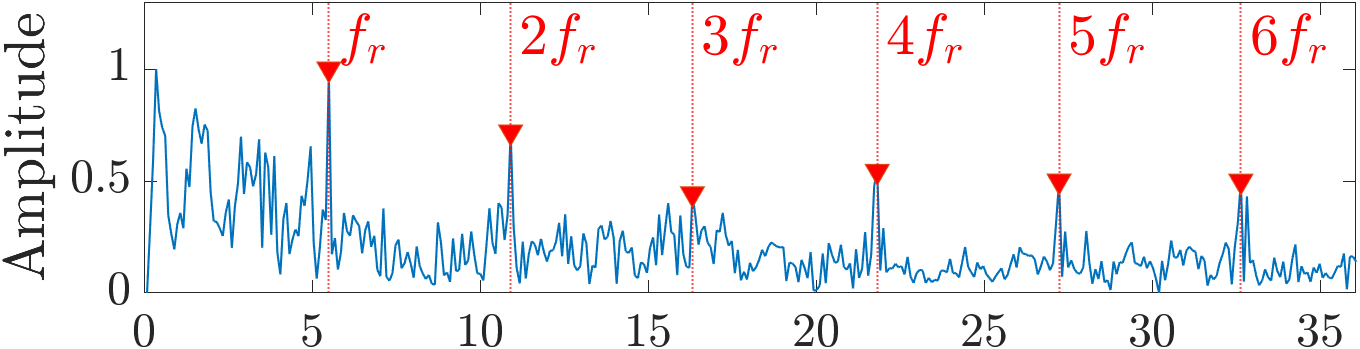}
    \end{subfigure}

    \begin{subfigure}[b]{\linewidth}
        \caption{Envelope Spectrum - sparsogram, ENVSI: 0.083}
        \centering
        \includegraphics[scale=0.32]{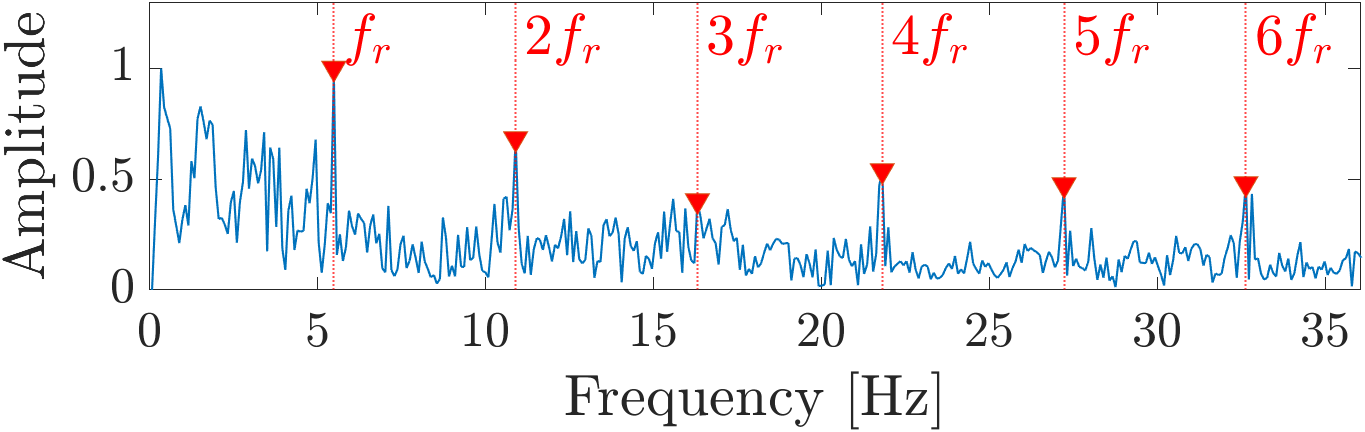}
    \end{subfigure}
    
    \caption{Envelope spectra of real signal with non-Gaussian noise filtered with analyzed methods.}
    \label{fig:envelope_spectra_nongauss}     
\end{figure}

\subsection{Robustness of SS-ONMF}
\label{subsub:robustness}

\begin{figure}[h!]
    \centering
    \includegraphics[scale=0.26]{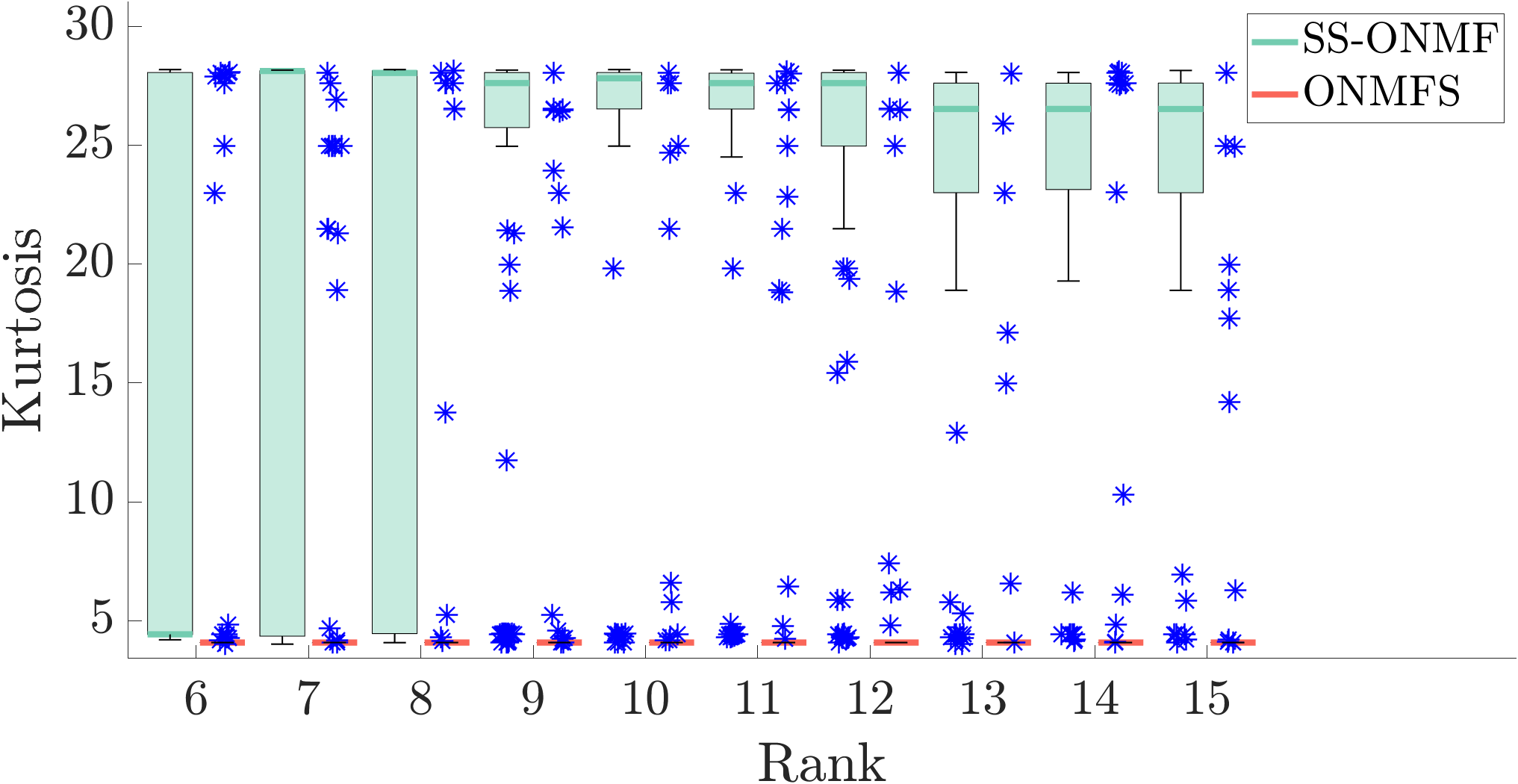}
    \caption{Boxplots of kurtosis measure obtained from the MC analysis of SS-ONMF and ONMFS, applied to the real Gaussian noisy signal versus ranks $6-15$.}
    \label{fig:vib1_kurtosis}
\end{figure}

\begin{figure}[h!]
    \centering
    \includegraphics[scale=0.26]{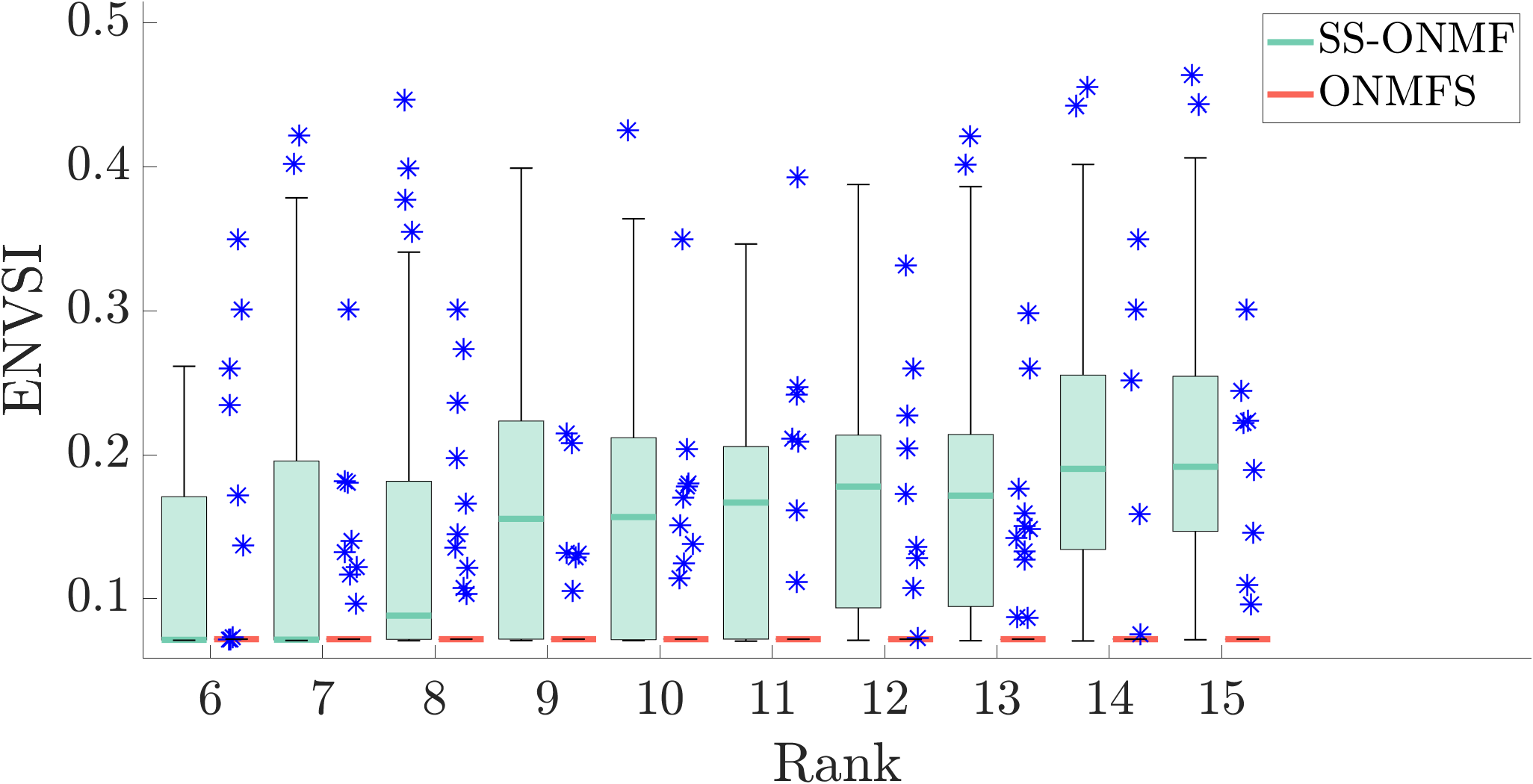}
    \caption{Boxplots of ENVSI measure obtained from the MC analysis of SS-ONMF and ONMFS, applied to the real non-Gaussian noisy signal versus ranks $6-15$.}
    \label{fig:dobry_envsis}
\end{figure}

Figure \ref{fig:vib1_kurtosis} illustrates the boxplots of the kurtosis values obtained from the MC analysis of SS-ONMF and ONMFS that were applied to the real Gaussian noisy signal versus the rank of approximation in the range $[6, 15]$. Similar results regarding the acoustic signal with non-Gaussian noise are depicted in Figure \ref{fig:dobry_envsis}. Both results (kurtosis and ENVSI measures) clearly demonstrate the superiority of SS-ONMF over ONMFS (which was discussed in Section \ref{subsub:real_results}), and SS-ONMF is much more robust to real signals. For the ranks greater than 9, the median values obtained with SS-ONMF are much higher than those obtained with ONMFS. Despite SS-ONMF is quite sensitive to initialization, the failure of SS-ONMF is highly unlikely for higher ranks. It is the opposite behavior to ONMFS which gives satisfactory results only for a few samples (outliers marked in black in both figures).



\subsection{Discussion}
The advantage of the proposed method is that it provides a very efficient filter that can be applied directly to the raw measured signal. It yields good results both in the time and frequency domains. The disadvantage of SS-ONMF is that it depends on stochastic sampling, which results in moderate sensitivity of a solution to initialization. However, due to the improved selection rules in SS-ONMF, its instability phenomena is considerably alleviated with reference to ONMFS. Moreover, to the best of our knowledge, there is no other approach to obtain hard orthogonality constraints in NMF. The proposed method was compared with various approaches, including NMF algorithms and the classical methods for fault detection. Among the NMF methods, ONMFS is very unstable, and only its outliers yield prominent results. The NMF-MU approach produces reasonable results, but the obtained filters are not as selective as in the case of SS-ONMF. Regarding the classical methods, SK is fast, easy to implement, and works well for signals with Gaussian noise. The drawback of SK is that it is sensitive to additional impulses (non-Gaussian), especially if the number of such impulses is large. CVB is designed for signals with non-Gaussian noise having large spiky impulses. Infogram is based on the Shannon entropy and was designed to extract repetitive transients from vibration signals. The drawback of the infogram is that it may have problems with heavy-tailed distributed noise.

However, the mentioned advantages and disadvantages of the compared methods (especially SK, CVB, and infogram) may not be so obvious for real signals. The presented results are slightly different than expected for SK, CVB, and infogram. This may be related to the complex nature of the real signals (which do not fully satisfy the assumptions). For the real signal with Gaussian noise, CVB gives better results than SK, and SK outperforms CVB for the non-Gaussian cases (see values of ENVSI). The differences may come from the properties of the analyzed signal itself (level of noise, number of impulses, etc.), because these methods are based on statistical properties of signals. The analysis of this phenomenon is beyond the scope of this study and we used these methods according to the guidelines in the original papers. Nevertheless, the important advantage of the proposed SS-ONMF method is that it works very well regardless of the signal properties for both Gaussian and non-Gaussian cases.


\section{Conclusions}
\label{sec:conclusions}

This study has presented a novel approach to vibration/acoustic fault detection in rolling element bearings. It is based on filter characteristic identification by factorizing the spectrogram with a new and efficient version of ONMF, where orthogonality constraints are imposed on the frequency profiles, and the stochastic sampling is used to explore the subspace of orthogonal and non-negative solutions.
The presented results highlight the importance of orthogonality constraints in obtaining a very selective filter characteristic that covers only the informative frequency band. Outside the band, the characteristic values are mostly zeros. The proposed approach provides better results than the state-of-the-art methods, e.g., such as the NMF-MU approach, and much better results than other methods for OFB selection -- SK, CVB,  infogram, {\color{black} and sparsogram.} For all the analyzed signals, the results confirm that the proposed approach is very effective both for the signals with Gaussian and non-Gaussion noise. 

The current study was limited to single-channel fault detection. Future studies should consider an extension of the proposed approach to multi-channel setting.
We can also address the development of new NMF algorithms for fault detection with additional constraints, such as sparsity, which can be more suitable for handling non-Gaussian disturbances. Finally, it is well known that the spectrogram is not the only TFR \cite{feng2013recent}. Using other TFRs, with better resolution in time and frequency, one should possibly obtain better results. We plan to take into account all these issues in our future papers.


\bibliography{bibliography}





\end{document}